\documentclass[12pt]{article}

\usepackage{amsmath,amsfonts,enumerate,array}
\allowdisplaybreaks
\usepackage{graphicx}
\usepackage[usenames]{color}
\usepackage[english]{babel}
\usepackage{fancybox}
\usepackage{chngpage} 

\usepackage{tikz-cd}

\newcommand{\captionfonts}{\small}

\makeatletter  
\long\def\@makecaption#1#2{%
  \vskip\abovecaptionskip
  \sbox\@tempboxa{{\captionfonts #1: #2}}%
 \ifdim \wd\@tempboxa >\hsize
    {\captionfonts #1: #2\par}
  \else
    \hbox to\hsize{\hfil\box\@tempboxa\hfil}%
  \fi
  \vskip\belowcaptionskip}
\makeatother   

\usepackage{mciteplus}

\usepackage[colorlinks=true,      linkcolor=blue,      urlcolor=blue,      filecolor=blue,      citecolor=blue,
      pdfstartview=FitH,     pdfpagemode=UseNone,      bookmarksopen=true]{hyperref}  
      
\usepackage[all]{hypcap}     

\usepackage{cite}

\begin{document}

\numberwithin{equation}{section}


\mathchardef\mhyphen="2D


\newcommand{\be}{\begin{equation}} 
\newcommand{\ee}{\end{equation}} 
\newcommand{\bea}{\begin{eqnarray}\displaystyle}
\newcommand{\eea}{\end{eqnarray}}
\newcommand{\bt}{\begin{tabular}}
\newcommand{\et}{\end{tabular}}
\newcommand{\bs}{\begin{split}}
\newcommand{\es}{\end{split}}

\newcommand{\I}{\text{I}}
\newcommand{\II}{\text{II}}

\renewcommand{\a}{\alpha}	
\renewcommand{\b}{\beta}
\newcommand{\g}{\gamma}		
\newcommand{\G}{\Gamma}
\renewcommand{\d}{\delta}
\newcommand{\D}{\Delta}
\renewcommand{\c}{\chi}			
\newcommand{\C}{\Chi}
\newcommand{\p}{\psi}			
\renewcommand{\P}{\Psi}
\newcommand{\s}{\sigma}		
\renewcommand{\S}{\Sigma}
\renewcommand{\t}{\tau}		
\newcommand{\e}{\epsilon}
\newcommand{\n}{\nu}
\newcommand{\m}{\mu}
\renewcommand{\r}{\rho}
\renewcommand{\l}{\lambda}

\newcommand{\nn}{\nonumber\\} 		
\newcommand{\newotimes}{}  				
\newcommand{\diff}{\,\text{d}}		
\newcommand{\h}{{1\over2}}				
\newcommand{\Gf}[1]{\G \Big{(} #1 \Big{)}}	
\newcommand{\floor}[1]{\left\lfloor #1 \right\rfloor}
\newcommand{\ceil}[1]{\left\lceil #1 \right\rceil}

\def\cA{{\cal A}} \def\cB{{\cal B}} \def\cC{{\cal C}}
\def\cD{{\cal D}} \def\cE{{\cal E}} \def\cF{{\cal F}}
\def\cG{{\cal G}} \def\cH{{\cal H}} \def\cI{{\cal I}}
\def\cJ{{\cal J}} \def\cK{{\cal K}} \def\cL{{\cal L}}
\def\cM{{\cal M}} \def\cN{{\cal N}} \def\cO{{\cal O}}
\def\cP{{\cal P}} \def\cQ{{\cal Q}} \def\cR{{\cal R}}
\def\cS{{\cal S}} \def\cT{{\cal T}} \def\cU{{\cal U}}
\def\cV{{\cal V}} \def\cW{{\cal W}} \def\cX{{\cal X}}
\def\cY{{\cal Y}} \def\cZ{{\cal Z}}

\def\mC{\mathbb{C}} \def\mP{\mathbb{P}}  
\def\mR{\mathbb{R}} \def\mZ{\mathbb{Z}} 
\def\mT{\mathbb{T}} \def\mN{\mathbb{N}}
\def\mH{\mathbb{H}} \def\mX{\mathbb{X}}
\def\CP{\mathbb{CP}}
\def\RP{\mathbb{RP}}
\def\Z{\mathbb{Z}}
\def\N{\mathbb{N}}
\def\H{\mathbb{H}}

\newcommand{\bin}[1]{{\bf {\color{blue} BG:}} {{\color{blue}\it#1}}}
\newcommand{\shaun}[1]{{\bf {\color{red} SH:}} {{\color{red}\it#1}}}


\addtolength{\skip\footins}{0pc minus 5pt}

\def\b{\bigskip}

\begin{flushright}
\end{flushright}
\vspace{15mm}
\begin{center}
{\LARGE Four-twist effects and monodromy \\\vspace{2mm} in symmetric orbifold CFTs}
\\
\vspace{18mm}
\textbf{Bin} \textbf{Guo}${}^{\dagger}$\footnote{guobin@phys.ethz.ch}~\textbf{and} ~ \textbf{Shaun}~ \textbf{D.}~ \textbf{Hampton}${}^{\ddagger}$\footnote{sdh2023@kias.re.kr}
\\
\vspace{8mm}
${}^{\dagger}$ \text{Institut f\"{u}r Theoretische Physik, ETH Z\"urich,}\\
	\hspace*{0.3cm} CH-8093 Z\"urich, Switzerland  \\
 \vspace{8mm}
 ${}^{\ddagger}$ \text{School of Physics},\\ Korea Institute for Advanced Study,\\ 85 Hoegiro, Dongdaemun-gu,\\ Seoul 02455, Korea
\vspace{3mm}
\end{center}

\vspace{2mm}

\thispagestyle{empty}

\begin{abstract}
\vspace{2mm}
Symmetric orbifold CFTs contain twist operators that can join and split copies of the CFT, leading to the creation of pairs from the vacuum. In this paper, we study the pair creation processes involving four twist-2 operators. In addition to the pair creation previously observed purely in the left or right moving sectors, we find a novel mixing between left and right movers during pair creation. This phenomenon arises from nontrivial monodromy conditions that originate from a genus-one covering surface, where left and right movers become coupled through the torus.

\end{abstract}
\newpage

\setcounter{page}{1}

\numberwithin{equation}{section} 

\tableofcontents

\newpage

\section{Introduction}

Symmetric orbifold CFTs have played a significant role in understanding the AdS$_3$/CFT$_2$ correspondence \cite{Maldacena:1997re, Strominger:1996sh, Seiberg:1999xz}. They provide known examples of classes of CFTs which admit a nice holographic dual. The tensionless limit of string theory \cite{Eberhardt:2018ouy,Eberhardt:2019ywk} corresponds to the free point of the symmetric orbifold CFT, while the supergravity limit corresponds to the strongly coupled regime of the CFT. At the free point \cite{Seiberg:1999xz,Larsen:1999uk,Dijkgraaf:1998gf,Jevicki:1998bm}, a marginal operator \cite{David:1999ec,Gomis:2002qi,Gava:2002xb} 
serves as a deformation, driving the CFT towards the supergravity regime. This marginal operator is built from a twist-2 operator dressed with certain supercharges. Understanding this deformation is crucial for interpolating between the tensionless and supergravity limits of the theory. A key first step is to study the effects of the twist-2 operator.  

Consider a seed CFT with target space  $M$; the symmetric orbifold CFT is constructed by taking $N$ copies of this seed CFT and orbifolding by the permutation group, resulting in a target space $ M^N/S_N $. Twisted sectors arise when some of the copies are joined into a longer copy while untwisted sectors correspond to configurations where all the copies remain unjoined.
The twist-2 operator is capable of joining and splitting copies of the CFT. Various effects of the twist-2 operator have been explored in the literature, such as in \cite{Avery:2010er,Avery:2010hs,Avery:2010qw,Burrington:2014yia,Carson:2014xwa,Carson:2014yxa,Guo:2022sos} for the case of a single twist-2 operator and \cite{Carson:2017byr,Carson:2015ohj, Carson:2016cjj} for the case of two twist-2 operators. 
During the joining and splitting process, the size of the vacuum changes. These changes produce effects that arise from Bogoliubov transformations of modes, a common procedure used to investigate physics when the vacuum undergoes changes, as in the case of quantum field theory in curved spacetime.

A significant effect produced by the action of the twist-2 operator is the creation of pairs of excitations from the vacuum. In previous studies involving one or two twist operators, pair creation was shown to occur independently in either the left-moving or right-moving sector, with other effects similarly manifesting in a decoupled manner between these sectors. When considering multiple twist operators, one might assume that the results for one or two twists could be sequentially applied during time evolution (with some UV regularization required). Hence, a natural assumption would be that the left and right movers remain decoupled in the effects of multiple twist operators. However, as we will show, this assumption does not hold.

In this paper, we investigate the simplest scenario in which the left- and right-moving sectors become coupled in the effect of multiple twist operators. Specifically, we consider the case of four twist operators acting on two untwisted copies of the CFT. The first twist joins the two untwisted copies into a single twist-2 copy, and the second twist splits them back into two separate copies. The third twist again joins them into a twist-2 copy, and the fourth twist separates them once more. Visualizing each copy as a circle, this sequence of twists creates a `string' loop between the second and third twist operators, which enables the coupling between the left- and right-moving sectors.

To explore this coupling, we use the covering map technique \cite{Lunin:2000yv,Lunin:2001pw}, a convenient method for calculating correlation functions involving twist operators. Twist operators introduce branch cuts, making fields multi-valued as they encircle the twist operators. By mapping to a covering space, these multi-valued fields become single-valued. In our setup with four twist operators, the covering space is a torus. On the torus, it is well known that the left- and right-moving sectors of free fields can couple through the nontrivial periodicity. This coupling on the covering space is ultimately transferred to the coupling of left- and right-moving sectors in the presence of the four twist operators. These nontrivial periodicities appear as monodromy conditions on the original base space. Monodromy constraints impose specific periodicity requirements on a field as it traverses closed loops around the twist operators \cite{Dixon:1986qv}. Since these constraints apply to the entire field including both the left- and right-moving sectors, they inherently couple the left and right movers. 

In this paper, we examine the pair creation effect induced by four twist-2 operators acting on two untwisted copies of the theory. The paper is organized as follows:
In Section \ref{effect of twist}, we review the symmetric orbifold CFT of a single free boson and the effects of twist operators. We also discuss the monodromy conditions and how they lead to the coupling of left- and right-moving sectors.
In Section \ref{correlation functions}, we compute the correlation functions of four twist operators and two bosonic fields using the covering map method. We show that the monodromy conditions manifest as periodicity constraints along the two cycles of the torus.
In Section \ref{pair creation}, we use the obtained correlation functions to compute the pair creation effects and analyze their behavior. Finally, in section \ref{results}, we collect our results, and in section \ref{discussion}, we discuss and conclude. 

\section{Effect of twist operators}\label{effect of twist}

\subsection{Symmetric product orbifold CFT}\label{orbifold review}

The symmetric product orbifold of $N$ copies of a seed CFT is described by the target space
\be
M^N/S_N
\ee 
where $M$ is the target space of the seed CFT and $S_N$ denotes the symmetric group.
In this paper, we will focus on the case where $N=2$ and $M = \mathbb{R}$, which corresponds to two copies of a free boson. We will denote these two copies as $X^{(1)}$ and $X^{(2)}$.

The base space is a cylinder parameterized by a complex coordinate $w$
\bea\label{cylinder coord}
w=\tau + i\sigma, \quad 
-\infty < \tau < \infty, \quad 0\leq \sigma < 2\pi
\eea
We will also use the complex plane coordinate $z$, which is related to the cylinder coordinate by 
\be
z=e^w
\ee

The action of $S_2$ permutes the two copies of the field into each other, leading to both untwisted and twisted sectors. In the untwisted sector, the fields $X^{(1)}$ and $X^{(2)}$ are periodic
\be
X^{(i)}(\tau,\sigma+2\pi)= X^{(i)}(\tau), \qquad   i =1,2
\ee
For each copy, we can define modes at constant $\tau$ as follows 
\begin{align} \label{alpha exp}
\a^{(i)}_{n}= \int_{\sigma=0}^{2\pi}{dw\over2\pi } e^{nw}\partial X^{(i)}, \qquad
\bar\a^{(i)}_{n}= \int_{\sigma=0}^{2\pi}{d\bar w\over2\pi } e^{n\bar w}
\bar\partial X^{(i)}
\end{align}
where $n$ is an integer. The vacuum $|0\rangle^{(i)}$ for copy $i$ is defined by
\bea
\a^{(i)}_{n}|0\rangle^{(i)}=0,\quad \bar\a^{(i)}_{n}|0\rangle^{(i)}=0,\quad n\geq0
\eea
The commutation relations are
\bea \label{comm relation singly wound}
[\a^{(i)}_{m},\a^{(j)}_{n}]=m\d^{ij}\d_{m+n,0},\quad [\bar\a^{(i)}_{m},\bar\a^{(j)}_{n}]=m\d^{ij}\d_{m+n,0},\quad [\a^{(i)}_{m},\bar\a^{(j)}_{n}]=0
\eea
which can be derived from the OPE
\begin{align} 
\partial X^{(i)}(z)\partial X^{(j)}(z')&\sim -{\d^{ij}\over (z-z')^2} + \text{regular terms}\nn
\bar \partial X^{(i)}(\bar z)\bar \partial X^{(j)}(\bar z')&\sim -{\d^{ij}\over (\bar z-\bar z')^2} + \text{regular terms}
\end{align}

There is also a twisted sector with the following boundary conditions
\be\label{twist bc}
X^{(1)}(\tau,\sigma+2\pi)= X^{(2)}(\tau), \qquad   X^{(2)}(\tau,\sigma+2\pi)= X^{(1)}(\tau)
\ee
To describe these two fields, it is convenient to introduce a single field defined on an extended base space with $0\leq \sigma<4\pi$ and periodicity $4\pi$. Since these fields and 
their mode expansions will not be used in this paper, we refer the reader to \cite{Avery:2010er,Carson:2014ena,Guo:2023czj} for more details.

In the $z$-plane, the ground state of this twisted sector corresponds to a local operator known as the twist operator, denoted by $\sigma_2$. In our context, this operator has conformal dimensions $h=\bar h =1/16$. Since we are focusing on this specific twist-2 operator, we will omit the subscript 2 for simplicity. 

For works on general twist operators and their correlations, please refer to \cite{Lunin:2000yv,Lunin:2001pw,Pakman:2009ab,Pakman:2009mi,Pakman:2009zz,Dei:2019iym}. The marginal deformation operator, which moves the D1D5 CFT from the free point towards the gravity regime, is constructed using the twist-2 operator. This deformation is essential for understanding anomalous dimensions\cite{Gava:2002xb,Gaberdiel:2015uca,Hampton:2018ygz,Guo:2020gxm,Benjamin:2021zkn,Lima:2021wrz,Apolo:2022fya,Guo:2022ifr,Gaberdiel:2023lco}, thermalization \cite{Hampton:2019csz,Hampton:2019hya}, and various aspects of black hole microstates\cite{Guo:2021ybz,Guo:2021gqd,Guo:2022and}.
Non-marginal deformations have also been investigated, such as the $T\bar T$ and $J\bar T$ deformations \cite{Chakraborty:2023wel}. Further work in orbifold CFTs have investigated the role of transport across boundaries and interfaces in these theories \cite{Baig_2023}.

\subsection{Monodromy}\label{monodromy sub}

Consider the $z$-plane with some insertions of the twist operator $\sigma$. 
We define the antisymmetric combination of the two fields, away from the twist operator's location, as 
\bea\label{antisymmetric X}
X \equiv {1\over\sqrt2}(X^{(1)} -  X^{(2)})
\eea
The field $X$ is antisymmetric under the interchange of the two copies. When tracing a loop around the twist operator, $X^{(1)}$ and $X^{(2)}$ are exchanged due to the boundary conditions (\ref{twist bc}). As a result, $X$ gets a minus sign upon a $2\pi$ rotation around the twist operator. Therefore, $X$ is only periodic under a $4\pi$ rotation around the twist operator.

The twist operator can be interpreted as generating a branch cut of order 2 on the $z$-plane, with $X^{(1)}$ and $X^{(2)}$ representing fields defined on the two Riemann sheets. The field $X$ becomes periodic when encircling a closed loop on the Riemann surface. More generally, $X$ is periodic\footnote{If the target space of the seed CFT includes identifications, the monodromy condition can shift by certain constants. In this paper, the target space of the seed CFT is taken to be $\mathbb{R}$, with no such identifications.} along any closed loop $C$ on the Riemann surface associated with the branched $z$-plane
\cite{Dixon:1986qv}
\bea\label{monodromy}
0=\Delta_{C} X= \oint_{C}dz\partial X + \oint_{C}d\bar z\bar{\partial} X
\eea 

The conditions in (\ref{monodromy}) impose nontrivial constraints that can couple the left and right movers. These constraints arise from nontrivial loops on the covering surface that cannot be contracted to a point. In the case of two twist operators, since the covering surface is a sphere, there are no nontrivial loops. Thus, no nontrivial monodromy constraints. The first nontrivial monodromy effect occurs with four twist operators, where the covering surface becomes a torus. In this paper, we focus on this effect, particularly the coupling between left and right movers, which has not yet been explored in the effect of twist operators.

\subsection{Bogoliubov ansatz for four twist effects} 

Consider the following problem: starting with the vacuum in the untwisted sector, we apply four twist operators and seek to determine the final state, which remains in the untwisted sector.
The vacuum in the untwisted sector is denoted by
\bea \label{initial state}
|0\rangle \equiv |0\rangle^{(1)}|0\rangle^{(2)}|\bar 0\rangle^{(1)}|\bar 0\rangle^{(2)}
\eea
where $|0\rangle^{(i)}|\bar 0\rangle^{(i)}$ is the vacuum state for the copy $i$, including both left and right movers. 
The state obtained by applying four twist operators can be described using the following Bogoliubov ansatz
\begin{align}\label{ansatz 4 first}
|\chi\rangle&\equiv\sigma(w_4,\bar w_4)\sigma(w_3,\bar w_3)\sigma(w_2,\bar w_2)\sigma(w_1,\bar w_1)|0\rangle\nn 
&=\mathcal C\exp\Big(\sum_{m,n>0}\gamma_{m,n}(w_i,\bar w_i)\a_{-m}\a_{-n} + \sum_{m,n>0}\beta_{m,n}(w_i,\bar w_i)\alpha_{-m}\bar{\alpha}_{-n}\nn
&\hspace{2cm}+\sum_{m,n>0}\bar\gamma_{m,n}(w_i,\bar w_i)\bar\a_{-m}\bar\a_{-n}\Big)|0\rangle
\end{align}
where 
\bea 
\mathcal C \equiv\langle 0| \sigma(w_4,\bar w_4)\sigma(w_3,\bar w_3)\sigma(w_2,\bar w_2)\sigma(w_1,\bar w_1)|0\rangle
\eea
We don't show the computation of $\mathcal C$ in this paper but it has been computed in several previous papers \cite{Dixon:1986qv,Lunin:2000yv}. Here, we focus only on the process of pair creation.
We prove the form of this ansatz in detail in appendix \ref{proof}.
The pair creation coefficients $\gamma_{m,n},\beta_{m,n},\bar\gamma_{m,n}$ depend on the positions of the twist operators $(w_i,\bar w_i)$. Since the twist operators are identical, these coefficients are symmetric with respect to the twist operator locations.
The bosonic modes in (\ref{ansatz 4 first}) are defined for $\tau>\tau_i$ in terms of the antisymmetric combination of bosonic fields $X$ (\ref{antisymmetric X}), 
\begin{align}\label{aabar}
    \alpha_n = {1\over2\pi}\int^{2\pi}_{\sigma=0}dw e^{nw}\partial X, \qquad
    \bar\alpha_n = {1\over2\pi}\int^{2\pi}_{\sigma=0}d\bar w e^{n\bar w}\bar\partial X
\end{align}
Using (\ref{alpha exp}), we obtain
\be \label{symm antisymm modes}
\alpha_n\equiv {1\over\sqrt2}(\a^{(1)}_n-\a^{(2)}_n),\quad \bar\alpha_n\equiv {1\over\sqrt2}(\bar\a^{(1)}_n-\bar\a^{(2)}_n)
\ee
Utilizing the commutation relations from (\ref{comm relation singly wound}), the commutation relations for $\alpha_n,\bar\alpha_n$ are
\begin{align}\label{aabar}
[\alpha_m,\alpha_n]=m\d_{m+n,0},\quad [\bar\alpha_m,\bar\alpha_n]=m\d_{m+n,0},\quad [\alpha_m,\bar\alpha_n]=0
\end{align}

Let's highlight some interesting features of the ansatz above. There is mixing between left and right movers characterized by $\beta_{m,n}$, preventing factorization between the two sectors. Additionally, the left-moving pair creation coefficient $\gamma_{m,n}(w_i,\bar w_i)$ is not a holomorphic function but instead depends on both $w_i$ and $\bar w_i$. These features distinguish it from the pair creation process with one or two twist operators \cite{Avery:2010er,Avery:2010hs,Burrington:2014yia,Carson:2014xwa,Carson:2014yxa,Carson:2017byr,Carson:2015ohj,Carson:2016cjj}, where there is no mixing between left and right movers: the left-moving pair creation is a holomorphic function, and the right-moving pair creation is an anti-holomorphic function. 

These \textit{novel} features arise due to the monodromy effects discussed in subsection \ref{monodromy sub}. As we will demonstrate later, the relation (\ref{monodromy}) for nontrivial loops leads to the mixing of left and right moving modes.
In the remainder of the paper, we will proceed by explicitly computing the pair creation coefficients $\gamma_{m,n}, \beta_{m,n}, \bar\gamma_{m,n}$.

\subsection{Left moving pair creation $\gamma_{m,n}$}

Using (\ref{ansatz 4 first}) and commutation relations (\ref{comm relation singly wound})
we have
\begin{align}\label{gamma11}
&\gamma_{m,n}
= {1\over 2mn}{\langle 0| \a_{n} \a_{m}\sigma(w_4,\bar w_4)\sigma(w_3,\bar w_3)\sigma(w_2,\bar w_2)\sigma(w_1,\bar w_1) |0\rangle\over \langle0| \sigma(w_4,\bar w_4)\sigma(w_3,\bar w_3)\sigma(w_2,\bar w_2)\sigma(w_1,\bar w_1)| 0\rangle}
\end{align}
with the factor of $2$ coming from the fact that $\gamma_{m,n}$ is symmetric in it's indices. The bosonic modes on the cylinder are 
\begin{equation}\label{boson cylinder}
    \alpha_m = {1\over2\pi}\int^{2\pi}_{\sigma=0,\tau>\tau_i}dw\, e^{mw}\, \partial_w X
\end{equation}
Let's map this amplitude to the $z$ plane through the map $z=e^{w}$.
Doing so we obtain the $z$-plane correlation function
\begin{align}\label{gamma11}
&\gamma_{m,n}
= {1\over 2mn}{\langle 0| \a_{n} \a_{m}\sigma(z_4,\bar z_4)\sigma(z_3,\bar z_3)\sigma(z_2,\bar z_2)\sigma(z_1,\bar z_1) |0\rangle\over \langle0| \sigma(z_4,\bar z_4)\sigma(z_3,\bar z_3)\sigma(z_2,\bar z_2)\sigma(z_1,\bar z_1)| 0\rangle}
\end{align}
We note that the Jacobian factors from the transformation of the four twist operators cancel out between the numerator and denominator. 
The bosonic modes (\ref{boson cylinder}) become 
\begin{equation}\label{bosonic modes z plane}
    \alpha_m = {1\over2\pi}\oint_{|z|>|z_i|}dz\, z^{m}\,\partial_z X
\end{equation}
Since these modes are defined in the final state, the range of it is $|z|>|z_i|$ thus yielding contours around $z=\infty$.
Therefore, inserting (\ref{bosonic modes z plane}) into (\ref{gamma11}) we obtain
\bea \label{pair creation z plane}
\gamma_{m,n}=-{1\over 2mn}\oint_{|z|>|z_i|}{dz\over2\pi }z^n\oint_{|z'|>|z_i|}{dz'\over2\pi }z'^m\,g(z,z';z_i,\bar z_i)
\eea
where the order of the  $z'$ and $z$ contours is unimportant, as both correspond to annihilation modes with $m,n>0$.
The correlation function, $g$, is defined as
\begin{align}\label{g}
g(z,z';z_i,\bar z_i) 
&\equiv-{\langle 0|\partial_z X\partial_{z'} X\sigma(z_4,\bar z_4)\sigma(z_3,\bar z_3)\sigma(z_2,\bar z_2)\sigma(z_1,\bar z_1) |0\rangle\over \langle 0| \sigma( z_4,\bar z_4)\sigma(z_3,\bar z_3)\sigma(z_2,\bar z_2)\sigma(z_1,\bar z_1)|0\rangle}
\end{align}
We will derive this expression by mapping this correlation function to the covering $t$-surface which is a genus one torus Riemann surface. We describe this in detail in the next section following closely \cite{Dixon:1986qv}.

\subsection{Left-right moving pair creation $\beta_{m,n}$}

Similarly the left-right moving pair creation coefficient $\beta_{m,n}$ is given by
\bea
\beta_{m,n} ={1\over mn}{\langle 0| \bar{\a}_{n} \a_{m}\sigma(w_4,\bar w_i)\sigma(w_3,\bar w_3)\sigma(w_2,\bar w_2)\sigma(w_1,\bar w_1) |0\rangle\over \langle 0| \sigma(w_4,\bar w_4)\sigma(w_3,\bar w_3)\sigma(w_2,\bar w_2)\sigma(w_1,\bar w_1)|0\rangle}
\eea
Mapping this correlation function to the $z$-plane gives 
\bea\label{z plane aabar}
\beta_{m,n} ={1\over mn}{\langle 0| \bar{\a}_{n} \a_{m}\sigma(z_4,\bar z_4)\sigma(z_3,\bar z_3)\sigma(z_2,\bar z_2)\sigma(z_1,\bar z_1) |0\rangle\over \langle 0| \sigma(z_4,\bar z_4)\sigma(z_3,\bar z_3)\sigma(z_2,\bar z_2)\sigma(z_1,\bar z_1)|0\rangle}
\eea
where the modes become
\begin{align}
       \alpha_m = {1\over2\pi}\oint_{|z|>|z_i|}dz z^{m}\partial_{z} X\ , \hspace{1cm}
    \bar\alpha_m = {1\over2\pi}\oint_{|\bar z|>|\bar z_i|}d \bar z\bar z^{m}\partial_{\bar z} X
\end{align}
Again, the contours are defined at $|z|>|z_i|$. 
Inserting these mode expansions into the $z$-plane correlation function (\ref{z plane aabar}) we obtain
\bea\label{beta}
\beta_{m,n} = -{1\over m n}\oint_{|\bar z|>|\bar z_i|}{d\bar z\over2\pi }\bar z^{n}\oint_{ |z'|>|z_i|}{dz'\over2\pi }z'^m\, b(\bar z, z';z_i,\bar z_i)
\eea
where the amplitude $b$ is defined as
\begin{align}\label{b}
b(\bar z, z';z_i,\bar z_i)&\equiv-{\langle0| \partial_{\bar z}X\partial_{ z'}X\sigma(z_4,\bar z_4)\sigma(z_3,\bar z_3)\sigma(z_2,\bar z_2)\sigma(z_1,\bar z_1) |0\rangle\over \langle0| \sigma(z_4,\bar z_4)\sigma(z_3,\bar z_3)\sigma(z_2,\bar z_2)\sigma(z_1,\bar z_1)|0\rangle}
\end{align}
In the next section, this expression will be derived by mapping it to a genus one covering surface.

\section{The covering map method}\label{correlation functions}

In this section, we derive the correlation functions (\ref{g}) and (\ref{b}) using the covering map method. To simplify the calculations, we first consider the locations of the twist operators at specific points $z_1=0, z_{2}=x,z_3=1,z_4=\infty$
\begin{align}\label{g and b}
g(z,z';x,\bar x) &\equiv -{\langle 0|\partial_z X\partial_{z'} X\sigma(\infty,\infty)\sigma(1, 1)\sigma(x,\bar x)\sigma(0,0) |0\rangle\over \langle 0| \sigma(\infty,\infty)\sigma(1, 1)\sigma(x,\bar x)\sigma(0,0)|0\rangle}
\nn 
\nn
b(\bar z, z';x,\bar x)&\equiv-{\langle0| \partial_{\bar z}X\partial_{ z'}X\sigma(\infty,\infty)\sigma(1,1)\sigma(x,\bar x)\sigma(0,0) |0\rangle\over \langle0| \sigma(\infty,\infty)\sigma(1,1)\sigma(x,\bar x)\sigma(0,0)|0\rangle}
\end{align}
Later, we will apply an SL$(2,\mathbb{C})$ transformation to map the twist operator locations from these specific points to arbitrary positions. 

\subsection{The covering map}

\begin{figure}
\centering
        \includegraphics[angle=270,width=8cm]{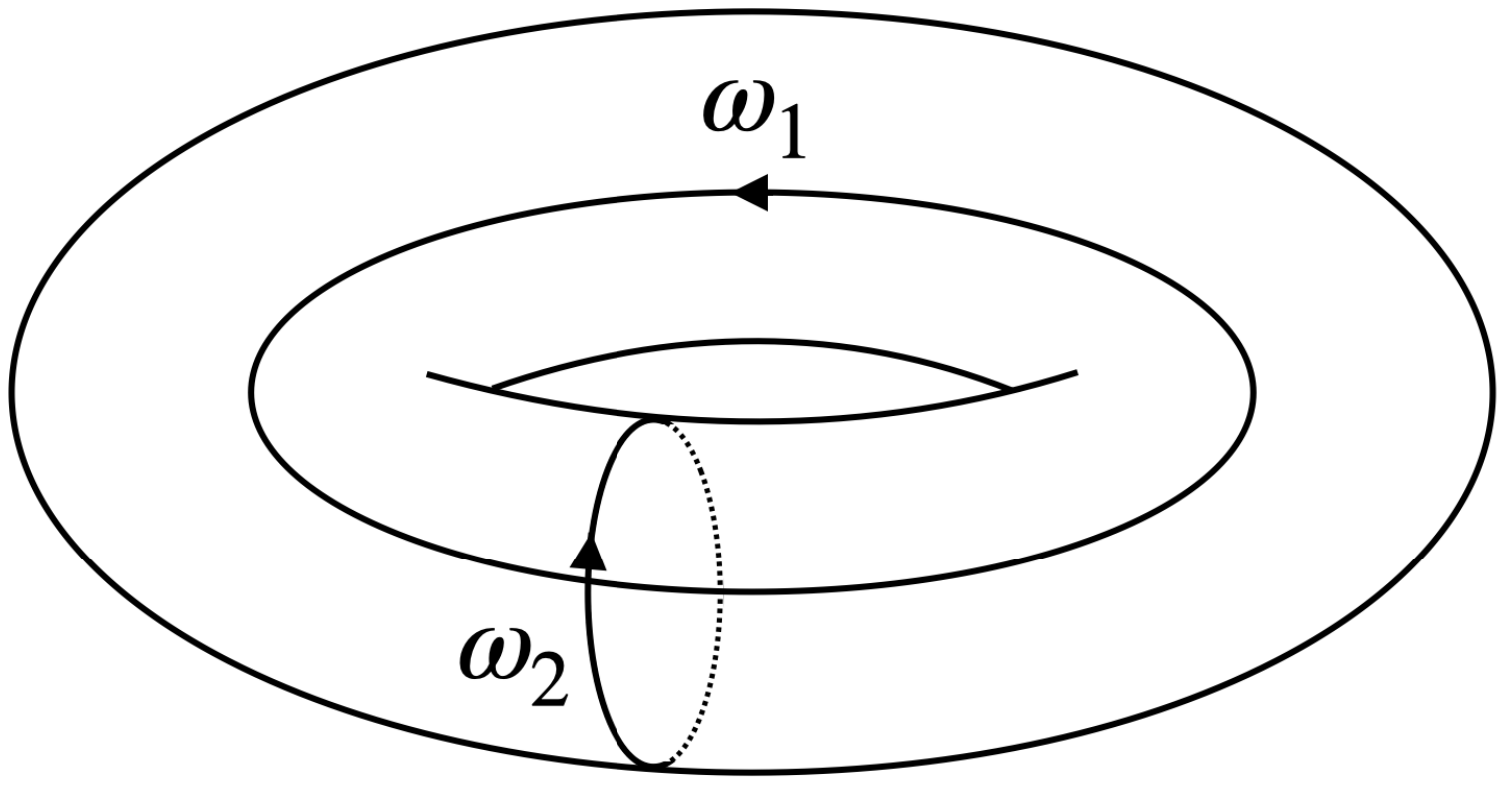}
\caption{The covering space, a genus one torus with cycles labeled by $\omega_i$.}\label{torus}
\end{figure}

In these correlators, the twist operators create branch points. To resolve these branch points, we map to a covering space characterized by the coordinate $t$. This covering map transforms the branched sphere into a double cover, which is a torus of genus $g=1$. The covering map is obtained by solving 
\bea\label{t plane transformation}
{dz\over dt}=C\big(z(z-x)(z-1)(z-z_{\infty})\big)^{1\over2}
\eea
where each branch point is of order $2$ and $C$ is a constant. In the limit $z_{\infty}\to\infty$, the solution to this equation gives the map 
\bea \label{t plane}
z = {\mathcal P(t) - e_1\over e_2 -e_1}
\eea
where $\mathcal P(t)$ is the Weierstrass $\mathcal P$ function, and the constants $e_1, e_2, e_3$ are defined as 
\be \label{half periods}
e_1= \mathcal P\big({1\over2}\big),\quad e_2= \mathcal P\big({\tau\over2}\big),\quad e_3 = \mathcal P\big({1\over2}+{\tau\over2}\big),\quad e_1 + e_2 + e_3 = 0
\ee
with $\tau$ being the modular parameter of the torus.
The covering space images of the twist locations on the plane can be read from the following values of $z(t)$
\be
z\big({1\over2}\big)=0,\ \quad z\big({\tau\over2}\big)=1,\ \quad z\big({1\over2}+{\tau\over2}\big)=x,\quad z(0)=\infty,\quad x = {e_3 - e_1\over e_2 - e_1}
\ee
The modular parameter $\tau$ is related to the cross ratio $x$ of the twist locations through the last relation, which can be written as
\be\label{cross ratio}
x = \bigg({\vartheta_4(\tau)\over\vartheta_3(\tau)}\bigg)^4
\ee
The fundamental domain of the torus is given by the coordinate identifications
\be
t \sim t + 1 \sim t+\tau
\ee
where these shifts define the periods of the torus, denoted by $\omega$
\begin{equation}\label{periods}
\omega=(\omega_1,\omega_2)=(1,\tau)
\end{equation}
as illustrated in figure \ref{torus}.
The Weierstrass $\mathcal P$ function is defined in terms of these periods as
\begin{equation}
\mathcal P(t)={1\over t^2} + \sum_{(m,n)\in\mathbb{Z}^2\backslash \{(0,0)\}}\bigg( {1\over (t - m\omega_1-n\omega_2)^2} - {1\over ( m\omega_1+n\omega_2)^2}\bigg)
\end{equation}
The function $\mathcal P$ is periodic under shifts by the torus periods
\bea\label{periodicity}
\mathcal P (t + \omega_i) = \mathcal P (t)
\eea

\subsection{Correlation functions on the covering surface}

We note that the Weierstrass $\mathcal P$ function is an even function, meaning $\mathcal P(t)=\mathcal P(-t)$. Therefore, a generic point $z$ on the sphere maps to two distinct points on the $t$-surface, each defining a different sheet on the $t$-surface. 
We recall that
\begin{equation}\label{def X}
\partial_z X = {1\over\sqrt2}(\partial_zX^{(1)}-\partial_zX^{(2)})
\end{equation}
Let us arbitrarily define sheet $1$ as corresponding to the coordinate $t$ and sheet $2$ to the coordinate $-t$. There is an inherent ambiguity in the sign of this definition. As we will see later, this sign ambiguity will appear naturally in the final result. 
Therefore the $z$-plane correlation functions (\ref{g and b}), map to the covering surface in the following way
\begin{align}\label{cf z}
    &g(z,z';x,\bar x)\nn
    =&-{1\over2}\left\langle\bigg(\Big({dz\over dt}\Big)^{-1}\partial_{t} X - \Big({dz\over d(-t)}\Big)^{-1} \partial_{-t}X\bigg)
    \bigg(\Big({dz'\over dt'}\Big)^{-1}\partial_{t'}X-\Big({dz'\over d(-t')}\Big)^{-1}\partial_{-t'}X\bigg)\right \rangle
    \nn
    =&-{1\over2}\Big({dz\over dt}\Big)^{-1}\Big({dz'\over dt'}\Big)^{-1}\Big(\langle\partial_{t} X \partial_{t'}X \rangle + \langle\partial_{-t} X \partial_{-t'} X\rangle+\langle\partial_{-t} X \partial_{t'}X \rangle  +\langle\partial_{t} X \partial_{-t'} X\rangle\Big)
   \nn
   \nn
  & b(\bar z,z';x,\bar x)\nn
    =&-{1\over2}\left\langle\bigg(\Big(\frac{d\bar z}{d\bar t}\Big)^{-1}\partial_{\bar t} X - \Big({d\bar z\over d(-\bar t)}\Big)^{-1} \partial_{-\bar t}X\bigg)
    \bigg(\Big({dz'\over dt'}\Big)^{-1}\partial_{t'}X-\Big({dz'\over d(-t')}\Big)^{-1}\partial_{-t'}X\bigg) \right\rangle
    \nn
    =&-{1\over2}\Big({d\bar z\over d\bar t}\Big)^{-1}\Big({dz'\over dt'}\Big)^{-1}\Big(\langle\partial_{\bar t} X \partial_{t'}X \rangle  +\langle\partial_{-\bar t} X \partial_{-t'} X\rangle+\langle\partial_{-\bar t} X \partial_{t'}X \rangle  +\langle\partial_{\bar t} X \partial_{-t'} X\rangle\Big)
\end{align}
where the two-point functions on the torus are well known and take the form 
\begin{align} \label{corr fun}
\langle\partial_{t} X \partial_{t'}X \rangle &= -\mathcal P(t-t')+c_1\nn
\langle\partial_{\bar t} X \partial_{t'}X \rangle&= c_2
\end{align}
where $c_1$ and $c_2$ are constants to be determined.
To obtain the first two-point function, note that it is periodic on the torus and exhibits a double pole at $t=t'$. This property uniquely identifies it as the Weierstrass $\mathcal P$ function up to an additive constant. For the second two-point function, since it is periodic on the torus without any poles, it must be constant. 

The two constants, $c_1$ and $c_2$, can be determined by using the periodicity conditions of the field $X$ on the torus, which give
\begin{align}\label{mon rel}
0 &= \int_0^1 dt\, \langle\partial_{t} X \partial_{t'}X \rangle+\int_0^1 d\bar t\, \langle\partial_{\bar t} X \partial_{t'}X \rangle\nn
0 &= \int_0^{\tau}dt\, \langle\partial_{t} X \partial_{t'}X \rangle+\int_0^{\tau}d\bar t\, \langle\partial_{\bar t} X \partial_{t'}X\rangle
\end{align}
Inserting the two-point functions (\ref{corr fun}) into these relations yield 
\begin{align}\label{c1 c2}
c_1&={1\over2i\text{Im}\tau}\bigg(\int_0^{\tau}dt \mathcal P(t)-\bar\tau\int_0^1 dt\mathcal P(t)\bigg) = \int_0^1dt \mathcal P(t) +{\pi\over\text{Im}\tau}
\nn  
c_2 &= -{1\over i\text{Im}\tau}\bigg(\int_0^{\tau} dt \mathcal P(t)-\tau\int_0^1 dt \mathcal P(t)\bigg)
 = -{2\pi\over\text{Im}\tau}
\end{align}
For the final equality for each constant, refer to Appendix \ref{der c1 c2}, where the integral $\int_0^1dt \mathcal P(t)$ is given explicitly in (\ref{intP}).

\subsection{Correlation functions on the $z$-plane}\label{Mapping to z plane}

Substituting (\ref{corr fun}) and (\ref{c1 c2}) into (\ref{cf z}), we obtain the $z$ plane correlation functions
\begin{align}\label{g and b z to t prime} 
g(z,z';x,\bar x)&=\bigg({dz\over dt}\bigg)^{-1}\bigg({dz'\over dt'}\bigg)^{-1}\bigg(\mathcal{P}(t-t') + \mathcal{P}(t+t') - 2\int_0^1dt\mathcal{P}(t)-{2\pi\over\text{Im}\tau}\bigg)
\nn 
b(\bar z,z';x,\bar x)&=\bigg({d\bar z\over d\bar t}\bigg)^{-1}\bigg({dz'\over dt'}\bigg)^{-1}{2\pi\over\text{Im}\tau}
\end{align}
These expressions are written on the covering space with the coordinate $t$. 
After performing the transformations to the $z$-plane, detailed in appendix \ref{mapping to z plane},
the above correlation functions can be rewritten as 
\begin{align}\label{g b sp pts}
g(z,z';x,\bar x)&=\, {{1\over2}z(z-1)(z'-x) + {1\over2}z'(z'-1)(z-x)+(z-z')^2 A(x,\bar x)\over z^{1\over2}(z-x)^{1\over2}(z-1)^{1\over2}z'^{{1\over2}}(z'-x)^{1\over2}(z'-1)^{1\over2}(z-z')^2}
\nn
b(\bar z,z';x,\bar x)&=\,{B(x,\bar x)\over\bar z^{1\over2}(\bar z - \bar x)^{1\over2}(\bar z - 1)^{1\over2} z'^{{1\over2}}( z' - x)^{1\over2}( z' - 1)^{1\over2}}
\end{align}
Notice that the above expressions have the correct expected branch points as indicated by their denominators. As the field circles a twist operator, the correlator changes sign, which reflects the ambiguity discussed after (\ref{def X}), where sheets $1$ and $2$ are interchanged.
The functions $A$ and $B$ are defined as
\begin{align} \label{AB}
A(x,\bar x) &= x(1-x){d\over dx}\ln(F(x)\bar F(1-\bar x)+F(1-x)\bar F(\bar x))
\nn 
B(x,\bar x)&={1\over\pi (F(x)\bar F(1-\bar x)+F(1-x)\bar F(\bar x))}
\end{align}
These functions can also be written in terms of elementary functions as
\begin{align} \label{EK}
A(x,\bar x) &= {(E(x) - (1-x)K(x))\bar K(1-\bar x)-( E(1-x) - xK(1-x) )\bar K(\bar x) \over2 (K(x)\bar K(1-\bar x)+  K(1-x)\bar K(\bar x))}
\nn 
B(x,\bar x) &= {\pi\over 4(K(x)\bar K(1-\bar x)+  K(1-x)\bar K(\bar x))}
\end{align}
where $K(y)$ is the complete elliptic integral of the first kind and $E(y)$ is the complete elliptic integral of the second kind. 
Writing the $A$ and $B$ in terms of $\tau$ using relations (\ref{more relations}), (\ref{im tau}) and (\ref{dtaudx}) yield
\begin{align} \label{A B tau}
A(\tau,\bar \tau)&= {i\over\pi\vartheta^4_3(\tau)}{d\over d\tau}\ln((\tau-\bar \tau)\vartheta_3^2(\tau)\bar{\vartheta}_3^2(\bar\tau)) 
\nn
B(\tau,\bar \tau)&=\bigg|{1\over2\pi \text{Im}\tau\,\vartheta^4_3(\tau)}\bigg|
\end{align}

An interesting feature of (\ref{g b sp pts}) is that $b$ is nonzero, indicating that
the holomorphic and antiholomorphic parts of the field are coupled due to the presence of four twist operators, whereas they remain decoupled for the case of two twist operators. This coupling can be traced back to the periodic conditions (\ref{mon rel}) imposed on nontrivial cycles of the covering $t$-torus. When these periodic conditions are transformed back to the $z$-plane, they take the form
\begin{align}\label{monodromy sp z plane}
0&=\oint_{C_{1}}dz\partial_z X + \oint_{C_{1}} d\bar z\partial_{\bar z} X
\nn
0&=\oint_{C_2}dz\partial_z X + \oint_{C_2}d\bar z\partial_{\bar z} X
\end{align}
Here, $C_{1}$ is a contour encircling either $z_1$ and $z_4$, or $z_2$ and $z_3$, which can be deformed into one another, making them equivalent. Similarly, $C_{2}$ is a contour surrounding either $z_1$ and $z_2$, or $z_3$ and $z_4$.
Both $C_{1}$ and $C_{2}$ are nontrivial cycles on the branched $z$-plane and together form a complete basis for all nontrivial cycles. These nontrivial cycles couple the holomorphic and antiholomorphic parts of the field. 

This coupling is expected to be a general feature in the presence of multiple twist operators (i.e., more than two), as it arises from the nontrivial cycles induced by the monodromy. In contrast, with only two twist operators, no such nontrivial cycles exist, as the cycle encircling the two twist operators can be deformed into a trivial cycle at infinity.

\subsection{General correlation functions from SL(2,$\mathbb C$)}

In this section, we apply an SL$(2,\mathbb{C})$ transformation to derive the correlation functions with twist operator locations at arbitrary points $z_1,z_2,z_3,z_4$ denoted by the coordinate $z$ from the correlation functions (\ref{g b sp pts}) with twist operator locations at special points $0,x,1,\infty$ denoted by the coordinate $\tilde z$. The conformal transformation is
\be\label{transformations}
\tilde z = {(z_1-z)(z_3-z_4)\over(z_1-z_3)(z-z_4)},\qquad
{d\tilde z\over d z}  
={(z_1-z_4)(z_4-z_3)\over (z_1-z_3)( z-z_4)^2}
\ee
The cross ratio $x$ is given by
\be\label{cross x}
x = {(z_1-z_2)(z_3-z_4)\over(z_1-z_3)(z_2-z_4)}
\ee
This conformal transformation gives the relation between correlation functions
\begin{align}
g(z,z';z_i,\bar z_i)&={d\tilde z\over dz}{d\tilde{z}'\over dz'} g(\tilde z,\tilde{z}';x,\bar x)
\nn 
b(\bar{z},z';z_i,\bar z_i)&={ d\bar{\tilde{z}}\over d\bar z}{ d\tilde{z}'\over dz'} b(\bar{\tilde{z}},\tilde{z}';x,\bar x)
\end{align}
Inserting the expressions (\ref{g b sp pts}), transformations (\ref{transformations}), and the cross ratio (\ref{cross x}) into the above relations yields expressions for the correlation functions at arbitrary twist operator locations
\begin{align}\label{g and b z plane}
&g(z,z';z_i,\bar z_i)\nn 
=&~~~ \, {1\over2}{(z-z_1)^{1\over2}(z-z_3)^{1\over2}(z'-z_2)^{1\over2}(z'-z_4)^{1\over2}\over(z-z_2)^{1\over2}(z-z_4)^{1\over2}(z'-z_1)^{1\over2}(z'-z_3)^{1\over2}}{1\over(z-z')^2}
\nn
&+{1\over2}{(z-z_2)^{1\over2}(z-z_4)^{1\over2}(z'-z_1)^{1\over2}(z'-z_3)^{1\over2}\over(z-z_1)^{1\over2}(z-z_3)^{1\over2}(z'-z_2)^{1\over2}(z'-z_4)^{1\over2}}{1\over(z-z')^2}
\nn
&+{A(z_i,\bar z_i)\over  (z-z_1)^{1\over2} (z-z_2)^{1\over2} (z-z_3)^{1\over2} (z-z_4)^{1\over2} (z'-z_1)^{1\over2} (z'-z_2)^{1\over2} (z'-z_3)^{1\over2} (z'-z_4)^{1\over2}}
\nn 
\nn
&b (\bar z,z';z_i,\bar z_i)\nn 
=&~{B(z_i,\bar z_i)\over  (\bar z-\bar z_1)^{1\over2} (\bar z-\bar z_2)^{1\over2} (\bar z-\bar z_3)^{1\over2} (\bar z-\bar z_4)^{1\over2} (z'-z_1)^{1\over2} (z'-z_2)^{1\over2} (z'-z_3)^{1\over2} (z'-z_4)^{1\over2}}
\end{align}
where $A(z_i,\bar z_i)$ and $B(z_i,\bar z_i)$ are given by  
\begin{align}\label{A B}
A(z_i,\bar z_i)&\equiv -(z_1-z_3)(z_2-z_4)A(x,\bar x)
\nn 
B(z_i,\bar z_i)&\equiv |z_1-z_3||z_2-z_4|B(x,\bar x)
\end{align}
where $A(x,\bar x)$ and $B(x,\bar x)$ are given by (\ref{AB}).
Notice that both $g$ and $b$ have the expected order 2 branch points at the location of the twist operators.

\subsection{Twist interchange symmetries}\label{interchange symm}

Since the four twist operators are identical, the correlation functions $g$ and $b$ (\ref{g and b z plane})
must be invariant under the interchange of the twist locations, up to a possible minus sign. This minus sign arises from the ambiguity mentioned earlier, where interchanging the locations of the twist operators may result in swapping copy 1 and copy 2 in (\ref{def X}). In this section, we show these symmetries explicitly. 

First, we consider the following interchanges, which correspond to the S transformation of the modular parameter $\tau$
\begin{align}\label{inter 1}
z_1  \leftrightarrow z_3\text{ or } z_2 & \leftrightarrow z_4, \qquad x\to 1-x = {(z_1-z_4)(z_2-z_3)\over(z_1-z_3)(z_2-z_4)},\qquad  \tau\to-{1\over\tau}
\end{align}
Under these transformations, and using (\ref{AB}) and (\ref{EK}),  we find that $A(x,\bar x)$ and $B(x,\bar x)$ transform as follows
\begin{align}\label{x1ox0}
    A(1-x,1-\bar x)
    &=-A(x,\bar x)
    \nn
    B(1-x,1-\bar x)&=B(x,\bar x)
\end{align}
Substituting these expressions into the general forms of $A(z_i,\bar z_i)$ and $B(z_i,\bar z_i)$ (\ref{A B}), we find
\begin{align}
    A(z_i,\bar z_i,z_1\leftrightarrow z_3) &= A(z_i,\bar z_i,z_2\leftrightarrow z_4) =A(z_i,\bar z_i)
    \nn 
    B(z_i,\bar z_i,z_1\leftrightarrow z_3) &= B(z_i,\bar z_i,z_2\leftrightarrow z_4) =  B(z_i,\bar z_i)
\end{align}
Thus, both $A(z_i,\bar z_i)$ and $B(z_i,\bar z_i)$ remain invariant under these interchanges. Therefore, the correlation functions $g$ and $b$ (\ref{g and b z plane}), are also invariant
\begin{align}
     g(z,z';z_i,\bar z_i,z_1\leftrightarrow z_3)&=g(z,z';z_i,\bar z_i,z_2\leftrightarrow z_4)= g(z,z';z_i,\bar z_i)
     \nn
     b(\bar z,z';z_i,\bar z_i,z_1\leftrightarrow z_3)&=b(\bar z,z';z_i,\bar z_i,z_2\leftrightarrow z_4) = b(\bar z,z';z_i,\bar z_i)
\end{align}

Next, we consider the following interchanges, which corresponds to the T transformation of the modular parameter $\tau$
\begin{align}\label{z1z4 transf}
z_1  \leftrightarrow z_4 \text{ or }z_2  \leftrightarrow z_3 ,\qquad x\to {1\over x} = {(z_1-z_3)(z_2-z_4)\over(z_1-z_2)(z_3-z_4)},\qquad \tau \to \tau +1
\end{align}
Applying these interchanges to the correlation functions $g$ and $b$ (\ref{g and b z plane}), we obtain
\begin{align}
&g(z,z';z_i,\bar z_i)- g(z,z';z_i,\bar z_i, z_2\leftrightarrow z_3)\nn [4pt]
=\ & { (z_1-z_4)(z_3-z_2)- 2 (z_1-z_3)(z_2-z_4) A(x,\bar x) + 2 (z_1-z_2)(z_3-z_4) A(\tfrac{1}{x},\tfrac{1}{\bar x}) \over2 (z-z_1)^{1\over2}(z-z_2)^{1\over2}(z-z_3)^{1\over2}(z-z_4)^{1\over2}(z'-z_1)^{1\over2}(z'-z_2)^{1\over2}(z'-z_3)^{1\over2}(z'-z_4)^{1\over2}}\nn [4pt]
=\ & {(z_1-z_3)(z_2-z_4)\big(x-1-2 A(x,\bar x) + 2 x A(\tfrac{1}{x},\tfrac{1}{\bar x}) \big)\over2 (z-z_1)^{1\over2}(z-z_2)^{1\over2}(z-z_3)^{1\over2}(z-z_4)^{1\over2}(z'-z_1)^{1\over2}(z'-z_2)^{1\over2}(z'-z_3)^{1\over2}(z'-z_4)^{1\over2}}\nn [10pt]
&b(\bar z,z';z_i,\bar z_i)- b(\bar z,z';z_i,\bar z_i, z_2\leftrightarrow z_3)\nn [4pt]
=\ &  {|z_1-z_3||z_2-z_4|B(x,\bar x)-|z_1-z_2||z_3-z_4|B(\tfrac{1}{x},\tfrac{1}{\bar x})\over(\bar z-\bar z_1)^{1\over2} (\bar z-\bar z_2)^{1\over2} (\bar z-\bar z_3)^{1\over2} (\bar z-\bar z_4)^{1\over2} (z'-z_1)^{1\over2} (z'-z_2)^{1\over2} (z'-z_3)^{1\over2} (z'-z_4)^{1\over2}}\nn [4pt]
=\ &  {|z_1-z_3||z_2-z_4|\big(B(x,\bar x)-|x|B(\tfrac{1}{x},\tfrac{1}{\bar x})\big)\over(\bar z-\bar z_1)^{1\over2} (\bar z-\bar z_2)^{1\over2} (\bar z-\bar z_3)^{1\over2} (\bar z-\bar z_4)^{1\over2} (z'-z_1)^{1\over2} (z'-z_2)^{1\over2} (z'-z_3)^{1\over2} (z'-z_4)^{1\over2}}
\end{align}
The above relations vanish due to the following relations 
\begin{align}\label{x1ox}
A(x,\bar x) &=xA(\tfrac{1}{ x},\tfrac{1}{ \bar x})  + {1\over2}(x-1)\nn 
B(x,\bar x) &= |x|B(\tfrac{1}{x},\tfrac{1}{\bar x})
\end{align}
Using these relations, we find that $g$ and $b$ are invariant under the interchanges
\begin{align}
g(z,z';z_i,\bar z_i, z_1\leftrightarrow z_4) &=g(z,z';z_i,\bar z_i, z_2\leftrightarrow z_3)=g(z,z';z_i,\bar z_i)\nn 
b(\bar z,z';z_i,\bar z_i, z_1\leftrightarrow z_4) &=b(\bar z,z';z_i,\bar z_i, z_2\leftrightarrow z_3)=b(\bar z,z';z_i,\bar z_i)
\end{align}
All other interchanges can be obtained from the above two interchanges given in (\ref{inter 1}) and (\ref{z1z4 transf}). For completeness, we now present the following interchanges
\begin{align}\label{z1z2}
z_1  \leftrightarrow z_2 \text{ or } z_3  \leftrightarrow z_4,\qquad x\to {x\over x-1} = {(z_2-z_1)(z_3-z_4)\over(z_1-z_4)(z_2-z_3)},\quad \tau\to {\tau\over1-\tau}
\end{align}
These interchanges can be obtained by successively applying (\ref{inter 1}), (\ref{z1z4 transf}) and (\ref{inter 1}), corresponding to an $STS$ transformation of the modular parameter $\tau$.
Applying these interchanges to the correlation functions $g$ and $b$, we find
\begin{align}
&g(z,z';z_i,\bar z_i)- g(z,z';z_i,\bar z_i, z_1\leftrightarrow z_2)\nn [4pt]
 =\ & { (z_1-z_2)(z_3-z_4)- 2 (z_1-z_3)(z_2-z_4) A(x,\bar x) + 2 (z_2-z_3)(z_1-z_4) A(\tfrac{x}{x-1},\tfrac{\bar x}{\bar x-1}) \over2 (z-z_1)^{1\over2}(z-z_2)^{1\over2}(z-z_3)^{1\over2}(z-z_4)^{1\over2}(z'-z_1)^{1\over2}(z'-z_2)^{1\over2}(z'-z_3)^{1\over2}(z'-z_4)^{1\over2}}\nn [4pt]
  =\ & {(z_1-z_2)(z_3-z_4)\big( 1-2\tfrac{1}{x} A(x,\bar x) + 2 \tfrac{1-x}{x} A(\tfrac{x}{x-1},\tfrac{\bar x}{\bar x-1}) \big)\over2 (z-z_1)^{1\over2}(z-z_2)^{1\over2}(z-z_3)^{1\over2}(z-z_4)^{1\over2}(z'-z_1)^{1\over2}(z'-z_2)^{1\over2}(z'-z_3)^{1\over2}(z'-z_4)^{1\over2}}\nn [10pt]
  &b(\bar z,z';z_i,\bar z_i)- b(\bar z,z';z_i,\bar z_i, z_1\leftrightarrow z_2)\nn [4pt]
  =\ & {|z_1-z_3||z_2-z_4|B(x,\bar x) - |z_1-z_4||z_2-z_3|B(\tfrac{x}{x-1},\tfrac{\bar x}{\bar x-1})\over(\bar z-\bar z_1)^{1\over2} (\bar z-\bar z_2)^{1\over2} (\bar z-\bar z_3)^{1\over2} (\bar z-\bar z_4)^{1\over2} (z'-z_1)^{1\over2} (z'-z_2)^{1\over2} (z'-z_3)^{1\over2} (z'-z_4)^{1\over2}}\nn [4pt]
  =\ & {|z_1-z_3||z_2-z_4|\big(B(x,\bar x) - |1-x|B(\tfrac{x}{x-1},\tfrac{\bar x}{\bar x-1})\big)\over(\bar z-\bar z_1)^{1\over2} (\bar z-\bar z_2)^{1\over2} (\bar z-\bar z_3)^{1\over2} (\bar z-\bar z_4)^{1\over2} (z'-z_1)^{1\over2} (z'-z_2)^{1\over2} (z'-z_3)^{1\over2} (z'-z_4)^{1\over2}}
\end{align}
Under these interchanges, we indeed find that $g$ and $b$ in (\ref{g and b z plane}) are invariant
\begin{align}
   g(z,z';z_i,\bar z_i,z_1\leftrightarrow z_2)&=g(z,z';z_i,\bar z_i,z_3\leftrightarrow z_4) = g(z,z';z_i,\bar z_i)\nn
   b(\bar z,z';z_i,\bar z_i,z_1\leftrightarrow z_2)&=b(\bar z,z';z_i,\bar z_i,z_3\leftrightarrow z_4) = b(\bar z,z';z_i,\bar z_i)
\end{align}
where we have used the following nontrivial relations
\begin{align}\label{xtoxoxm1}
A(x,\bar x) &= (1-x)A(\tfrac{x}{ x-1},\tfrac{\bar x}{ \bar x-1}) + {x\over 2}\nn
B(x,\bar x)&=  |1-x|B(\tfrac{x}{x-1},\tfrac{\bar x}{\bar x-1})
\end{align}
which can be derived from (\ref{x1ox0}) and (\ref{x1ox}).

\subsection{Reduction to two twist operators}\label{4 to 2}

In this subsection, we show that the correlation function involving four twist operators can be reduced to a correlation function involving two twist operators when any two of the four twist operators are brought together. Specifically, We will show this for the case when $z_3\to z_4$
 \begin{align}\label{g to two}
g(z,z';z_i,\bar z_i,z_3\to z_4)&=\lim_{z_3\to z_4}-{\langle 0|\partial_z X\partial_{z'} X\sigma(z_4,\bar z_4)\sigma(z_3,\bar z_3)\sigma(z_2,\bar z_2)\sigma(z_1,\bar z_1) |0\rangle\over \langle 0| \sigma( z_4,\bar z_4)\sigma(z_3,\bar z_3)\sigma(z_2,\bar z_2)\sigma(z_1,\bar z_1)|0\rangle}\nn
&=-{\langle 0|\partial_z X\partial_{z'} X\sigma(z_2,\bar z_2)\sigma(z_1,\bar z_1) |0\rangle\over \langle 0| \sigma(z_2,\bar z_2)\sigma(z_1,\bar z_1)|0\rangle}
\end{align}
Due to the interchange symmetry of twist operators, as discussed in the previous section, similar relations hold when any two twist operators are brought together.

The expression in the last line of (\ref{g to two}) is given by
\begin{align}\label{two twist corr}
    -{\langle 0|\partial_z X\partial_{z'} X\sigma(z_2,\bar z_2)\sigma(z_1,\bar z_1) |0\rangle\over \langle 0| \sigma(z_2,\bar z_2)\sigma(z_1,\bar z_1)|0\rangle} ={(z-z_1)(z'-z_2) + (z-z_2)(z'-z_1)  \over2 (z-z_1)^{1\over2}(z-z_2)^{1\over2}(z'-z_1)^{1\over2}(z'-z_2)^{1\over2}(z-z')^2}
\end{align}
This result can be obtained from the following analytic behavior. As $z$ approaches $z_i$, it behaves as $(z-z_i)^{-1/2}$ due to the presence of twist operators. As $z$ approaches $z'$, it behaves as $(z-z')^{-2}$ with additional regular terms but without a single pole. When $z\to \infty$, it decays as $z^{-2}$ since the operator $\partial X$ has dimension 1. Notice that this $z\to \infty$ condition rules out the possibility of a nontrivial monodromy term in the numerator, such as $(z-z')^2 A(z_i,\bar z_i)$.

Let us take the limit $z_3\to z_4$ in $g$. Using (\ref{EK}), we find
\be
    x = {(z_1-z_2)(z_3-z_4)\over(z_1-z_3)(z_2-z_4)}\to0\implies A(x,\bar x)\to0
\ee
Thus, it is straightforward to verify that $g$, as found in (\ref{g and b z plane}), satisfies the above condition (\ref{g to two}).

We have therefore shown that when any two twist operators coincide, the correlation function involving four twist operators reduces to a correlation function involving two twist operators. We have also numerically verified that the pair creation coefficients collected in section \ref{results} exhibit the same behavior.

\section{Pair creation for four twists}\label{pair creation}

In this section, using the results of section \ref{Mapping to z plane} we derive pair creation coefficients for four twist operators. These expressions consist of three contributions: one for the left-moving pair, $\gamma_{m,n}$, one for the right-moving pair, $\bar{\gamma}_{m,n}$ and one for the left-right mixed pair, $\beta_{m,n}$. 
Since computing $\bar{\gamma}_{m,n}$ is analogous to computing $\gamma_{m,n}$ we compute the latter explicitly and then note that the former is obtained by replacing the relevant quantities with their complex conjugates. 

\subsection{Computing $\gamma_{m,n}$}
 
We compute the pair creation process related to modes $\alpha_{-m}\alpha_{-n}$.
Recalling equation (\ref{pair creation z plane}), pair creation, $\gamma_{m,n}$,  is given by, 
\bea 
\gamma_{m,n}=-{1\over 2mn} \oint_{|z|>|z_i|}{dz\over 2\pi }z^n\oint_{|z'|>|z_i|}{dz'\over2\pi }z'^mg(z,z';z_i,\bar z_i)
\eea
where though the order of integration is not important, we make an explicit choice in order to compute the contour integrals.
Inserting (\ref{g and b z plane}) 
and splitting up each term yields
\begin{equation}\label{I II III}
    \gamma_{m,n}\equiv\gamma_{\text{I}} + \gamma_{\text{II}} + \gamma_{\text{III}}
\end{equation}
where
\begin{align}\label{I II III pr}
\gamma_{\text{I}}&\equiv  -{1\over 4mn}\oint_{|z|>|z_i|}{dz\over2\pi } z^n\oint_{|z'|>|z_i|}{dz'\over 2\pi }  z'^m ~{(z-z_1)^{1\over2}(z-z_3)^{1\over2}(z'-z_2)^{1\over2}(z'-z_4)^{1\over2}\over(z-z_2)^{1\over2}(z-z_4)^{1\over2}(z'-z_1)^{1\over2}(z'-z_3)^{1\over2}}{1\over(z-z')^2}\nn
\nn
\gamma_{\text{II}}&\equiv- {1\over 4mn}\oint_{|z|>|z_i|}{dz\over2\pi } z^n\oint_{|z'|>|z_i|}{dz'\over2\pi }  z'^m{(z-z_2)^{1\over2}(z-z_4)^{1\over2}(z'-z_1)^{1\over2}(z'-z_3)^{1\over2}\over(z-z_1)^{1\over2}(z-z_3)^{1\over2}(z'-z_2)^{1\over2}(z'-z_4)^{1\over2}}{1\over(z-z')^2}\nn
\nn
\gamma_{\text{III}}&\equiv {1\over 2mn}A(x,\bar x)(z_1-z_3)(z_2-z_4)\oint_{|z|>|z_i|}{dz\over2\pi } z^n\oint_{|z'|>|z_i|}{dz'\over2\pi }  z'^m\nn 
&\quad{1\over  (z-z_1)^{1\over2} (z-z_2)^{1\over2} (z-z_3)^{1\over2} (z-z_4)^{1\over2} (z'-z_1)^{1\over2} (z'-z_2)^{1\over2} (z'-z_3)^{1\over2} (z'-z_4)^{1\over2}}
\end{align}
We note that the contours defined above are to be evaluated at the singularities at $z=\infty$ since we can freely deform them to this location on the plane.
We also note that terms $\gamma_{\text{I}}$ and $\gamma_{\text{II}}$ are related in the following way
\bea
\gamma_{\text{II}} = \gamma_{\text{I}}(z_1\leftrightarrow z_2,z_3\leftrightarrow z_4)
\eea
Let's proceed to compute the contour integrals for each of the three terms. To do so we will need the following expansions around $z\sim\infty$ in which $|z|\gg |z_i|$
\be\label{expansion}
(z-z_i)^{\pm{1\over2}}=z^{\pm{1\over2}}(1-z_{i}z^{-1})^{\pm{1\over2}}
=\sum_{k\geq0}{}^{\pm{1\over2}}C_{k}(-1)^kz^k_{i}z^{-k\pm{1\over2}}
\ee
where ${}^{p}C_q$ is the Binomial coefficient
\be
{}^{p}C_q={p!\over q! (p-q)! }
\ee
Next we proceed to compute the three terms in (\ref{I II III pr}).

\paragraph{Term I and II}

Inserting the expansion (\ref{expansion}) into term I in (\ref{I II III pr}) yields 
\begin{align}\label{gamma I}
\gamma_{\text{I}} =& -{1\over 4mn}\sum_{k_1,k'_1,k_2,k'_2,k_3,k'_3,k_4,k'_4\geq0}{}^{{1\over2}}C_{k_1}{}^{-{1\over2}}C_{k_2}{}^{{1\over2}}C_{k_3}{}^{-{1\over2}}C_{k_4}{}^{-{1\over2}}C_{k'_1}{}^{{1\over2}}C_{k'_2}{}^{-{1\over2}}C_{k'_3}{}^{{1\over2}}C_{k'_4}\nn 
&(-1)^{k_1+k_2+k_3+k_4+k'_1+k'_2+k'_3+k'_4}z^{k_1+k'_1}_{1}z^{k_2+k'_2}_{2}z^{k_3+k'_3}_{3}z^{k_4+k'_4}_{4}
\nn
&\oint_{|z|>|z_i|}{dz\over 2\pi } z^{n-(k_1+k_2+k_3+k_4)}\oint_{|z'|>|z_i|}{dz'\over2\pi }z'^{m-(k'_1+k'_2+k'_3+k'_4)}~{1\over(z-z')^2}
\end{align}
To perform the integration, we apply a coordinate transformation that maps $\infty$ to $0$
\begin{equation}\label{z to u}
z={1\over u},\quad z'={1\over u'}
\end{equation}
Inserting these transformations into the last line of (\ref{gamma I}), we obtain
\begin{align}\label{gamma I p}
\oint_{|u'|>|u|}{du'\over2\pi }  u'^{-m+k'_1+k'_2+k'_3+k'_4} \oint_{u=0}{du\over 2\pi } u^{-n+k_1+k_2+k_3+k_4} ~{1\over(u-u')^2}
\end{align}
where we pick a particular order of the contour where $|u'|>|u|$.
The overall sign stays the same because the minus sign coming from the differentials is compensated by the minus sign arising from the contours changing direction.  
Computing first the $u$ integral at $u=0$ gives
\begin{align}\label{gamma I p}
i\oint_{u'=0}{du'\over2\pi }  u'^{-n-m+k'_1+k'_2+k'_3+k'_4-1}\big(n-(k_1+k_2+k_3+k_4)\big)\bigg|_{n-(k_1+k_2+k_3+k_4)>0}
\end{align}
where the constraint is a necessary condition to yield a nonzero result. Computing the $u'$ integral at $u'=0$ yields
\begin{align} \label{I}
\big(- n + k_1 + k_2 + k_3 + k_4\big)\d_{m+n-(k_1+k'_1+k_2+k'_2+k_3+k'_3+k_4+k'_4),0}\big|_{n-(k_1+k_2+k_3+k_4)>0}
\end{align}
The constraint imposed by Kronecker delta is 
\bea \label{k4p}
k'_4=m+n-(k_1+k_2+k_3+k_4+k'_1+k'_2+k'_3)
\eea
Since $k_i, k'_j\geq 0$ for $i,j=1,2,3,4$, from (\ref{k4p}) and the constraint in (\ref{I}) we deduce the following constraints on the sums
\begin{align} \label{constr 1}
k'_3&\leq m+n - (k_1+k_2+k_3+k_4+k'_1+k'_2)\nonumber\\
k'_2&\leq m+n - (k_1+k_2+k_3+k_4+k'_1)\nonumber\\
k'_1&\leq m+n - (k_1+k_2+k_3+k_4)\nonumber\\
k_4&\leq n-(k_1+k_2+k_3+1)\nonumber\\
k_3&\leq n-(k_1+k_2+1)\nonumber\\
k_2&\leq n-(k_1+1)\nonumber\\
k_1&\leq n-1
\end{align} 
Imposing constraints (\ref{constr 1}) we get
\bea \label{I}
\gamma_{\text{I}}&=&{1\over 4mn}(-1)^{m+n}\sum_{k_1=0}^{n-1}~\sum_{k_2=0}^{n-(k_1+1)}~\sum_{k_3=0}^{n-(k_1+k_2+1)}~\sum_{k_4=0}^{n-(k_1+k_2+k_3+1)}~\sum_{k'_1=0}^{m+n - (k_1+k_2+k_3+k_4)}\nn 
&&~\sum_{k'_2=0}^{m+n - (k_1+k_2+k_3+k_4+k'_1)}\sum_{k'_3=0}^{m+n - (k_1+k_2+k_3+k_4+k'_1+k'_2)}\big(n - (k_1 + k_2 + k_3 + k_4)\big)\nn 
&&{}^{{1\over2}}C_{k_1}{}^{-{1\over2}}C_{k_2}{}^{{1\over2}}C_{k_3}{}^{-{1\over2}}C_{k_4}{}^{-{1\over2}}C_{k'_1}{}^{{1\over2}}C_{k'_2}{}^{-{1\over2}}C_{k'_3}{}^{{1\over2}}C_{m+n-(k_1+k'_1+k_2+k'_2+k_3+k'_3+k_4)}\nn 
&&z^{k_1+k'_1}_{1}z^{k_2+k'_2}_{2}z^{k_3+k'_3}_{3}z^{m+n-(k_1+k'_1+k_2+k'_2+k_3+k'_3)}_{4}
\eea
For $\gamma_{\text{II}}$, we simply make the exchanging $z_1\leftrightarrow z_2$ and $z_3\leftrightarrow z_4$ in $\gamma_{\text{I}}$. 

\paragraph{Term III} For $\gamma_{\text{III}}$, using expansion (\ref{expansion}), we have
\bea\label{Term 3}
\gamma_{\text{III}}&=&A(x,\bar x)(z_1-z_3)(z_2-z_4)\nn 
&&\bigg({1\over \sqrt2n}\sum_{k_1,k_2,k_3,k_4\geq0}{}^{-{1\over2}}C_{k_1}{}^{-{1\over2}}C_{k_2}{}^{-{1\over2}}C_{k_3}{}^{-{1\over2}}C_{k_4}(-1)^{k_1+k_2+k_3+k_4}z_1^{k_1}z_2^{k_2}z_3^{k_3}z_4^{k_4}\nn 
&&\quad \oint_{z=\infty}{dz\over2\pi } z^{n-(k_1+k_2+k_3+k_4+2)}\bigg)\nn
&&\bigg({1\over \sqrt2m}\sum_{k'_1,k'_2,k'_3,k'_4\geq0}{}^{-{1\over2}}C_{k'_1}{}^{-{1\over2}}C_{k'_2}{}^{-{1\over2}}C_{k'_3}{}^{-{1\over2}}C_{k'_4}(-1)^{k'_1+k'_2+k'_3+k'_4}z_1^{k'_1}z_2^{k'_2}z_3^{k'_3}z_4^{k'_4}\nn 
&&\quad \oint_{z'=\infty}{dz'\over2\pi }  z'^{m-(k'_1+k'_2+k'_3+k'_4+2)}\bigg) 
\eea
Again making the transformation (\ref{z to u}) and computing residues at $u=0$ yields
\begin{align}
\gamma_{\text{III}}&=-A(x,\bar x)(z_1-z_3)(z_2-z_4)\nn 
&\quad\bigg({1\over \sqrt2n}\sum_{k_1,k_2,k_3,k_4\geq0}{}^{-{1\over2}}C_{k_1}{}^{-{1\over2}}C_{k_2}{}^{-{1\over2}}C_{k_3}{}^{-{1\over2}}C_{k_4}(-1)^{k_1+k_2+k_3+k_4}z_1^{k_1}z_2^{k_2}z_3^{k_3}z_4^{k_4}\nn 
&\qquad \d_{n-(k_1+k_2+k_3+k_4+1),0}\bigg)\nn
&\quad\bigg({1\over \sqrt2m}\sum_{k'_1,k'_2,k'_3,k'_4\geq0}{}^{-{1\over2}}C_{k'_1}{}^{-{1\over2}}C_{k'_2}{}^{-{1\over2}}C_{k'_3}{}^{-{1\over2}}C_{k'_4}(-1)^{k'_1+k'_2+k'_3+k'_4}z_1^{k'_1}z_2^{k'_2}z_3^{k'_3}z_4^{k'_4}\nn
&\qquad \d_{m-(k'_1+k'_2+k'_3+k'_4+1),0}\bigg)
\end{align}
Imposing conditions from the kronecker delta terms and using similar arguments to place restrictions on the remaining sums yields
\begin{align}\label{III}
\gamma_{\text{III}}=&-A(x,\bar x)(z_1-z_3)(z_2-z_4)\nn 
&\bigg({1\over \sqrt2m}(-1)^{m}\sum_{k'_1=0}^{m-1}~\sum_{k'_2=0}^{m-(k'_1+1)}~\sum_{k'_3=0}^{m-(k'_1+k'_2+1)}{}^{-{1\over2}}C_{k'_1}{}^{-{1\over2}}C_{k'_2}{}^{-{1\over2}}C_{k'_3}{}^{-{1\over2}}C_{m-(k'_1+k'_2+k'_3+1)}\nn
&\qquad z_1^{k'_1}z_2^{k'_2}z_3^{k'_3}z_4^{m-(k'_1+k'_2+k'_3+1)}\bigg)
\nn
&\bigg({1\over \sqrt2n}(-1)^{n}\sum_{k_1=0}^{n-1}~\sum_{k_2=0}^{n-(k_1+1)}~\sum_{k_3=0}^{n-(k_1+k_2+1)}{}^{-{1\over2}}C_{k_1}{}^{-{1\over2}}C_{k_2}{}^{-{1\over2}}C_{k_3}{}^{-{1\over2}}C_{n-(k_1+k_2+k_3+1)}\nn
&\qquad z_1^{k_1}z_2^{k_2}z_3^{k_3}z_4^{n-(k_1+k_2+k_3+1)}\bigg)
\end{align}
Collecting together the three contributions, the final expression for $\gamma_{m,n}$ in (\ref{I II III}) then becomes
\begin{align} \label{gamma final}
\gamma_{m,n} 
=\ & {1\over 4mn}(-1)^{m+n}\sum_{k_1=0}^{n-1}~\sum_{k_2=0}^{n-(k_1+1)}~\sum_{k_3=0}^{n-(k_1+k_2+1)}~\sum_{k_4=0}^{n-(k_1+k_2+k_3+1)}~\sum_{k'_1=0}^{m+n - (k_1+k_2+k_3+k_4)}\nn 
&~\sum_{k'_2=0}^{m+n - (k_1+k_2+k_3+k_4+k'_1)}\sum_{k'_3=0}^{m+n - (k_1+k_2+k_3+k_4+k'_1+k'_2)}\big(n - (k_1 + k_2 + k_3 + k_4)\big)\nn 
&{}^{{1\over2}}C_{k_1}{}^{-{1\over2}}C_{k_2}{}^{{1\over2}}C_{k_3}{}^{-{1\over2}}C_{k_4}{}^{-{1\over2}}C_{k'_1}{}^{{1\over2}}C_{k'_2}{}^{-{1\over2}}C_{k'_3}{}^{{1\over2}}C_{m+n-(k_1+k'_1+k_2+k'_2+k_3+k'_3+k_4)}\nn 
&\times\Big(z^{k_1+k'_1}_{1}z^{k_2+k'_2}_{2}z^{k_3+k'_3}_{3}z^{m+n-(k_1+k'_1+k_2+k'_2+k_3+k'_3)}_{4}\nn
&\qquad + z^{k_1+k'_1}_{2}z^{k_2+k'_2}_{1}z^{k_3+k'_3}_{4}z^{m+n-(k_1+k'_1+k_2+k'_2+k_3+k'_3)}_{3}\Big) 
\nn [3pt]
& \! - A(x,\bar x)( z_1- z_3)( z_2 - z_4)\nn 
&\quad\times\bigg({1\over \sqrt2m}(-1)^{m}\sum_{k'_1=0}^{m-1}~\sum_{k'_2=0}^{m-(k'_1+1)}~\sum_{k'_3=0}^{m-(k'_1+k'_2+1)}{}^{-{1\over2}}C_{k'_1}{}^{-{1\over2}}C_{k'_2}{}^{-{1\over2}}C_{k'_3}{}^{-{1\over2}}C_{m-(k'_1+k'_2+k'_3+1)}
\nn
&\qquad\quad z_1^{k'_1}z_2^{k'_2}z_3^{k'_3}z_4^{m-(k'_1+k'_2+k'_3+1)}\bigg)
\nn
&\quad\times\bigg({1\over \sqrt2n}(-1)^{n}\sum_{k_1=0}^{n-1}~\sum_{k_2=0}^{n-(k_1+1)}~\sum_{k_3=0}^{n-(k_1+k_2+1)}{}^{-{1\over2}}C_{k_1}{}^{-{1\over2}}C_{k_2}{}^{-{1\over2}}C_{k_3}{}^{-{1\over2}}C_{n-(k_1+k_2+k_3+1)}
\nn
&\qquad\quad z_1^{k_1}z_2^{k_2}z_3^{k_3}z_4^{n-(k_1+k_2+k_3+1)}\bigg) 
\end{align}

\subsection{Computing $\beta_{m,n}$}

We compute the pair creation process for the modes $\alpha_{-m}\bar{\alpha}_{-n}$ characterized by $\beta_{m,n}$. We start with (\ref{beta})
\bea 
\beta_{m,n} = -{1\over mn}\oint_{|\bar z|>|\bar z_i|}{d\bar z\over2\pi }\bar z^{n}\oint_{ |z'|>|z_i|}{dz'\over2\pi }z'^mb(\bar z, z';z_i,\bar z_i)
\eea
where the full expression for $b(\bar z, z';z_i,\bar z_i)$ is given in (\ref{g and b z plane}). We see that computing $\beta_{m,n}$ is analogous to computing term III of $\gamma_{m,n}$, (\ref{Term 3}), with the appropriate modifications. We therefore don't show the explicit computation but only the final expression which is given to be
\begin{align}\label{betamn}
\beta_{m,n} &= B(x,\bar x)|z_1-z_3||z_2-z_4|
\nn
&\quad\times\bigg({1\over m}(-1)^{m}\sum_{k'_1=0}^{m-1}~\sum_{k'_2=0}^{m-(k'_1+1)}~\sum_{k'_3=0}^{m-(k'_1+k'_2+1)}{}^{-{1\over2}}C_{k'_1}{}^{-{1\over2}}C_{k'_2}{}^{-{1\over2}}C'_{k_3}{}^{-{1\over2}}C_{m-(k'_1+k'_2+k'_3+1)}
\nn
&\qquad\quad z_1^{k'_1}z_2^{k'_2}z_3^{k'_3}z_4^{m-(k'_1+k'_2+k'_3+1)}\bigg)
\nn 
&\quad\times\bigg({1\over n}(-1)^{n}\sum_{k_1=0}^{n-1}~\sum_{k_2=0}^{n-(k_1+1)}~\sum_{k_3=0}^{n-(k_1+k_2+1)}{}^{-{1\over2}}C_{k_1}{}^{-{1\over2}}C_{k_2}{}^{-{1\over2}}C_{k_3}{}^{-{1\over2}}C_{n-(k_1+k_2+k_3+1)}
\nn
&\qquad\quad \bar z_1^{k_1}\bar z_2^{k_2}\bar z_3^{k_3}\bar z_4^{n-(k_1+k_2+k_3+1)}\bigg)
\end{align}

\subsection{Numerical results}\label{sec numeri}

In this subsection, we numerically analyze the pair creation coefficients, working in the Lorentzian signature on the cylinder. 

The cylinder coordinates $w$ and $\bar w$ are defined as follows
\begin{equation}\label{zibarzi}
 z=e^{w},\quad \bar z=e^{\bar w}
\end{equation}
and 
\begin{equation}\label{wibarwieuc}
 w = \tau + i\sigma,\quad \bar w =\tau - i \sigma
\end{equation}
Performing a Wick rotation to the Lorentzian signature, where $\tau=it$,  the $w$'s become
\begin{equation}\label{mink sig}
    w = i(t + \sigma),\quad \bar w =i(t -  \sigma)
\end{equation}
The operation of absolute value appearing in various places in the above expressions is understood as an action in Euclidean signature. For a function $f$, the absolute value is defined as
\begin{equation}
|f(z_i,\bar z_i)|=\sqrt{f(z_i,\bar z_i)\bar f( \bar z_i, z_i)}
\end{equation}
where the coordinates $z$ and $\bar z$ are given by (\ref{zibarzi}) and (\ref{mink sig}).

We first observe that $\gamma_{m,n}$ ($\bar\gamma_{m,n}$) and $\beta_{m,n}$ are invariant under any set of twist interchanges $z_i\leftrightarrow z_j$ ($w_i\leftrightarrow w_j$) for $i,j=1,2,3,4$. This symmetry arises from the fact that the correlation functions from which they are derived, $g$ and $b$, are invariant under such twist interchanges, as demonstrated in subsection \ref{interchange symm}. Furthermore, we numerically verify that the $\gamma_{m,n}$ (\ref{gamma final}) and $\beta_{m,n}$ (\ref{betamn}) exhibit such symmetry.

By direct inspection of the expressions (\ref{gamma final}) and (\ref{betamn}), as well as through numerical testing, we find that the pair creation coefficients $\gamma_{m,n}$ and $\beta_{m,n}$ have a periodicity of $2\pi$ for any configuration of twist locations
\be
\gamma_{m,n}(t_i+2\pi)=\gamma_{m,n}(t_i), \qquad \beta_{m,n}(t_i+2\pi)=\beta_{m,n}(t_i)
\ee

\begin{figure}
\centering
        \includegraphics[width=7.5cm]{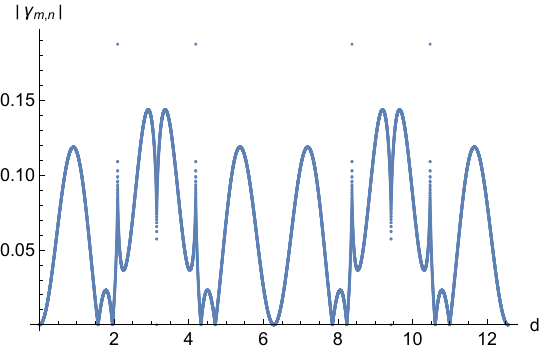}   
        \includegraphics[width=7.5cm]{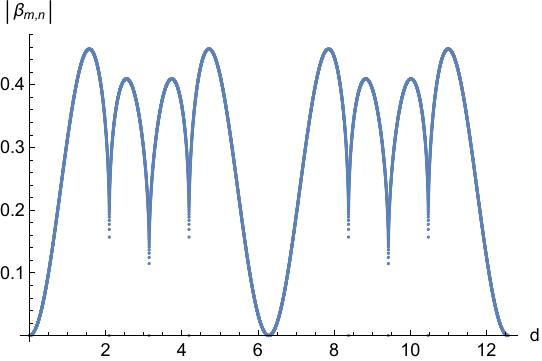}
\caption{$|\gamma_{m,n}|$ (left) and $|\beta_{m,n}|$ (right) vs. distance $d$ between equally spaced twist operators in time, with $\sigma_1=\sigma_2=\sigma_3=\sigma_4=0$, and $t_1=0$, $t_2=d$, $t_3=2d$, $t_4=3d$ with a range of $d$ from $0$ to $4\pi$ for $m=n=1$. 
}\label{gamma_beta_4pi_m1}
\end{figure}

In the following, we consider the configuration where all twist operators are equally spaced in time, with the separation denoted as $d$, and sitting at the same spatial location, which we set to $\sigma_i=0$.
Due to the twist interchange symmetry and the periodicity, the plot exhibits reflective symmetry around $d=\pi$ and periodicity of $2\pi$, as clearly shown in Fig. \ref{gamma_beta_4pi_m1}.

\begin{figure}
\centering
        
        \includegraphics[width=8cm]{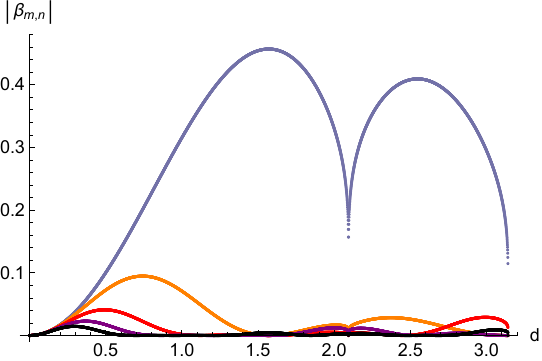}
\caption{ $|\beta_{m,n}|$ vs. distance $d$ between equally spaced twist operators in time for $m=n=1$ (blue), $2$ (orange), $3$ (red), $4$ (purple), and $5$ (black), with $\sigma_1=\sigma_2=\sigma_3=\sigma_4=0,$ and $t_1=0$, $t_2=d$, $t_3=2d$, $t_4=3d$, for a range of $d$ from $0$ to $\pi$. }\label{fig2Dmd}
\end{figure}

Fig. \ref{fig2Dmd} shows the behavior of $|\beta_{m,n}|$ as a function of the equal time separation, $d$, between consecutive twist operators, for different values of $m=n$, over $0\leq d\leq\pi$. It is evident that $\beta_{m,n}$ takes the largest value at $d={\pi\over2}$ which is the largest separation configuration amongst the four twist operators within a periodic $2\pi$ interval, and for the lowest mode number $m=n=1$. 

This behavior aligns with the general understanding of the role of twist operators. Twist operators can cut and join the copies of the CFTs. In our scenario, with time ordering, the first and third twist operators join two untwisted copies of the CFT into a twisted copy, while the second and fourth twist operators split the twisted copy back into two untwisted copies. As a result, consecutive twist operators tend to cancel each other's effects when they are close. 
Therefore, the maximum effect occurs when the twist operators are most widely separated, which corresponds to the configuration with $d=\frac{\pi}{2}$.

For higher energy modes, where the wavelength is shorter, these modes are primarily produced through smaller separations of twist operators. The smaller separation leads to a stronger cancellation between the effects of the twist operators, resulting in a smaller pair creation coefficient $\beta_{m,n}$.
  
We note that $\beta_{m,n}$ vanishes not only at $d=0$, where the twist operators are not separated at all, but also for $d={2\pi\over3}$ and for $d=\pi$. At $d={2\pi\over3}$, the first and fourth twist operators are separated by a distance of $2\pi$, which gives the same effect as if they were not separated due to the $2\pi$ periodicity, thus yielding a configuration which is equivalent to that of two twist operators, where $\beta_{m,n}$ is zero. At $d=\pi$, the twist configuration is such that the first and third twist operators, as well as the second and fourth twist operators, are separated by $2\pi$. Again, periodicity implies that this is equivalent to a configuration where the first and third twist operators coincide, as well as the second and fourth twist operators.
This explains the second maximum that occurs between $d={2\pi\over3}$ and $d=\pi$. For higher mode numbers, similar features appear with additional local maxima and zeros, as the number of terms in $\beta_{m,n}$ (\ref{betamn}) increases, yielding a more complicated zero-maximum structure, as illustrated in Fig. \ref{fig2Dmd}. 

\begin{figure}
\centering
        \includegraphics[width=7.4cm]{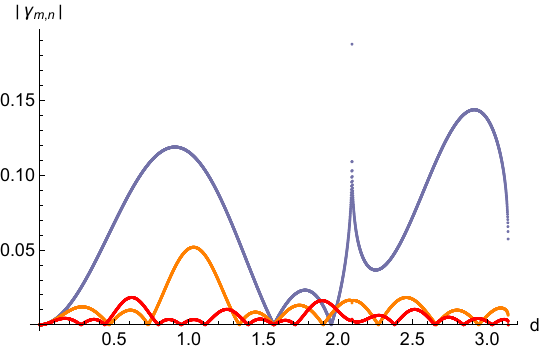}
        \includegraphics[width=7.4cm]{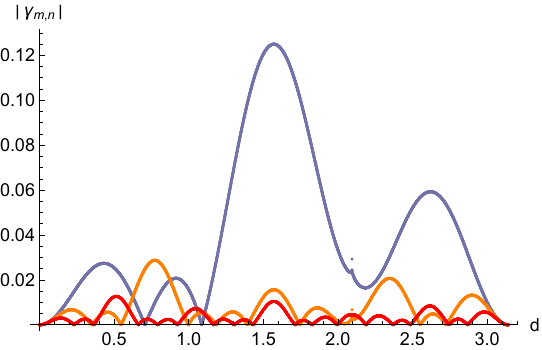}
\caption{$|\gamma_{m,n}|$ with $m=n=1$ (blue), $3$ (orange), $5$ (red) (left), and $m=n=2$ (blue), $4$ (orange), $6$ (red) (\text{right}) vs. distance $d$ between equally spaced twist operators in time with $\sigma_1=\sigma_2=\sigma_3=\sigma_4=0,$ and $t_1=0$, $t_2=d$, $t_3=2d$, $t_4=3d$ with a range of $d$ from $0$ to $\pi$.}\label{gammafig2Dmd}
\end{figure}

Figure \ref{gammafig2Dmd} is a plot of $|\gamma_{m,n}|$ as a function of the time separation $d$ between successive twist operators. The left plot corresponds to odd mode numbers $m=n=1,3,5$ and the right plot corresponds to even mode numbers $m=n=2,4,6$. We distinguish odd and even cases since their behavior is slightly different, where for odd mode numbers, pair creation vanishes at $d={\pi\over2}$, whereas for even mode numbers, pair creation reaches a local maximum.
As the mode numbers increase, the number of oscillations also increase, while the amplitudes decrease due to the greater cancellation between twist operators, as previously discussed. 
Additionally, $\gamma_{m,n}$ vanishes at $d=0$ and $d=\pi$ for the same reasons as $\beta_{m,n}$, but does not vanish at $d={2\pi\over3}$, where $\gamma_{m,n}$ yields the two twist result.

\begin{figure}
\centering
        \includegraphics[width=8.0cm]{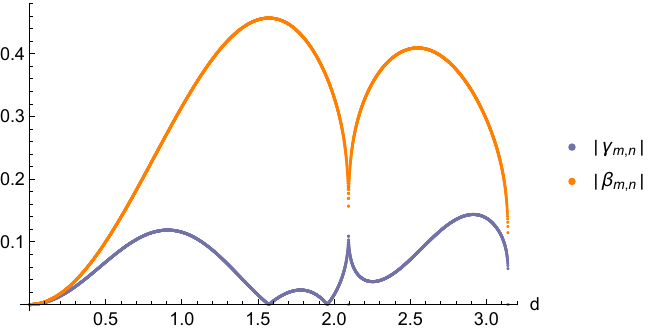}
        \includegraphics[width=8.0cm]{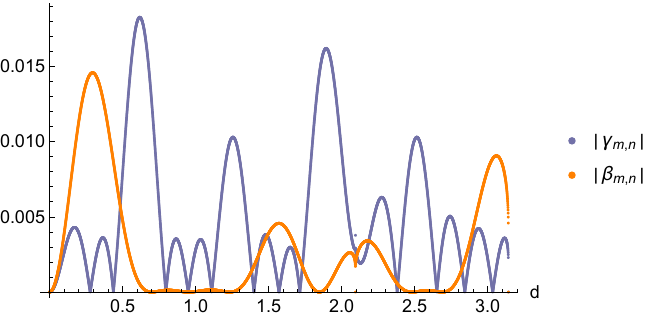}
\caption{$|\gamma_{m,n}|$ (blue) and $|\beta_{m,n}|$ (orange) vs. distance $d$ between equally spaced twist operators in time, with $\sigma_1=\sigma_2=\sigma_3=\sigma_4=0,$ and $t_1=0$, $t_2=d$, $t_3=2d$, $t_4=3d$ with a range of $d$ from $0$ to $\pi$ for $m=n=1$ (left) and $m=n=5$ (right).}\label{gamma_betam}
\end{figure}

The comparison between the left-right mixed pair creation $\beta_{m,n}$ and the usual unmixed pair creation $\gamma_{m,n}$ is captured nicely in the left plot of Fig. \ref{gamma_betam}. 
For $m=n=1$, at equal twist separations of $d={\pi\over2}$ (the maximum equal separation within a periodic $2\pi$ interval), $\beta_{m,n}$ (orange) reaches its maximum value. In contrast, $\gamma_{m,n}$ (blue) vanishes for this configuration. Thus we see a configuration in which mixed pair creation clearly dominates over unmixed pair creation. This pattern holds for any odd mode number $m=n$.

\begin{figure}
\centering
        \includegraphics[width=7.4cm]{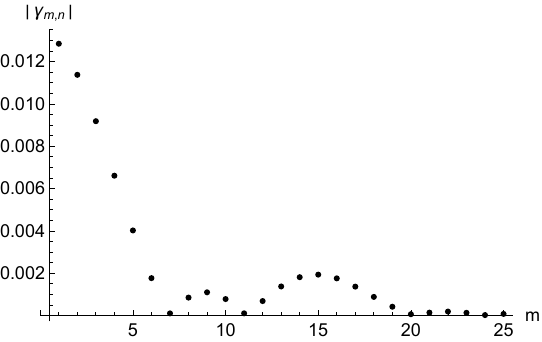}
        \includegraphics[width=7.4cm]{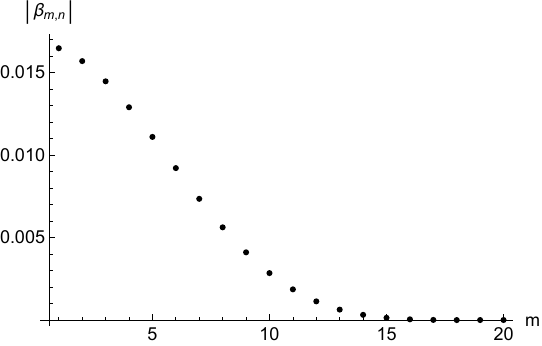}
\caption{$|\gamma_{m,n}|$ (\text{left}), $|\beta_{m,n}|$ (\text{right}) vs. $m$ for $m=n$  and  $\sigma_1=\sigma_2=\sigma_3=\sigma_4=0,$ and $t_1=0$, $t_2={\pi\over16}$, $t_3={\pi\over8}$, $t_4={3\pi\over16}$.}\label{fig1}
\end{figure}

More generally, we observe a key distinction between $\gamma_{m,n}$ and $\beta_{m,n}$ in various plots, such as the right plot in Fig. \ref{gamma_betam}. The unmixed pair creation $\gamma_{m,n}$ generally displays both global (nontrivial monodromy) and local behavior, characterized by oscillatory motion throughout the interval. In contrast, the left-right mixed pair creation, $\beta_{m,n}$, although exhibiting some oscillations, has fewer distinct peaks, implying that its behavior is less sensitive to twist location details. To explore this further, Fig. \ref{fig1} presents the case for small separations $d = \frac{\pi}{16}$.

We see that $\gamma_{m,n}$ has a maximum at the lowest energies with a local maximum at $m=15$ and with an envelop which continues to decay. This local maximum represents the fact the $\gamma_{m,n}$ is sensitive to the small twist separation which is approximately the inverse of the energy. 
Similar behavior is also observed in the simpler case of two twists. 
However, $\beta_{m,n}$ shows no local maximum but instead decays to zero smoothly, suggesting that $\beta_{m,n}$ is significantly less sensitive to the details of the twist separations, especially for small separations relative to the spatial size of $2\pi$, and lacks the oscillatory features seen in $\gamma_{m,n}$.

\begin{figure}
\centering
        \includegraphics[width=7.6cm]{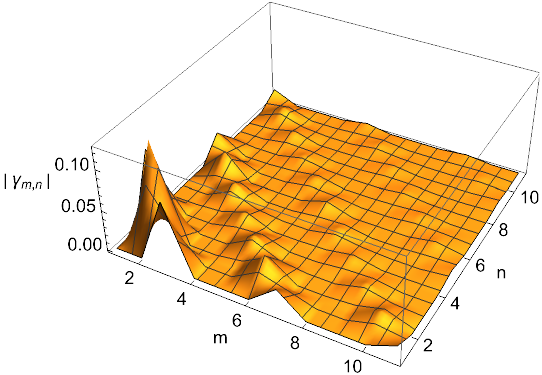}
        \includegraphics[width=7.6cm]{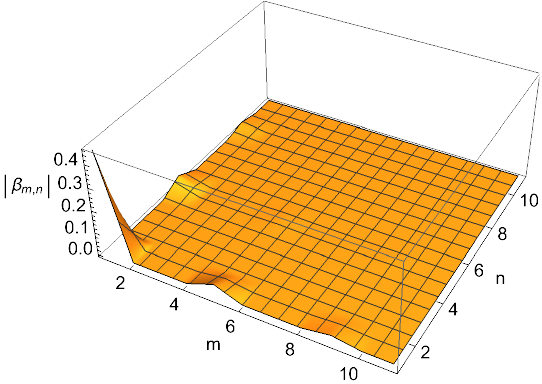}
\caption{$|\gamma_{m,n}|$ (left), $|\beta_{m,n}|$ (right)  vs. $m,n$  for  $\sigma_1=\sigma_2=\sigma_3=\sigma_4=0,$ and $t_1=0$, $t_2={\pi\over2}$, $t_3=\pi$, $t_4={3\pi\over2}$. Although $m,n$ are integers, smooth interpolation is used for better visualization.}\label{fig3D}
\end{figure}

In Fig. \ref{fig3D}, we plot $\gamma_{m,n}$ and $\beta_{m,n}$ for arbitrary values of $m,n$ for the separation $d=\frac{\pi}{2}$, which is the largest separation between four twist operators in a periodic $2\pi$ interval. Although $m,n$ are integers, we include smooth interpolation for better visualization. 
Both $\gamma_{m,n}$ and $\beta_{m,n}$ are symmetric between $m$ and $n$. The symmetry in $\gamma_{m,n}$ follows from the fact that $\alpha_{m}\alpha_{n}$ and $\alpha_{n}\alpha_{m}$ represent the same modes. The symmetry in $\beta_{m,n}$ arises from the configuration where all twist operators are located at the same spatial point, $\sigma=0$, which has a $\sigma\to -\sigma$ symmetry that interchanges left and right movers. This symmetry would break in more general configurations with different spatial locations.

\section{Collecting Results}\label{results}

Here we collect all of the relevant results for pair creation in the presence of four twist operators. 
The left-moving pair creation coefficient is given by
\begin{align} \label{gamma final prime}
\gamma_{m,n} 
=\ & {1\over 4mn}(-1)^{m+n}\sum_{k_1=0}^{n-1}~\sum_{k_2=0}^{n-(k_1+1)}~\sum_{k_3=0}^{n-(k_1+k_2+1)}~\sum_{k_4=0}^{n-(k_1+k_2+k_3+1)}~\sum_{k'_1=0}^{m+n - (k_1+k_2+k_3+k_4)}\nn 
&~\sum_{k'_2=0}^{m+n - (k_1+k_2+k_3+k_4+k'_1)}\sum_{k'_3=0}^{m+n - (k_1+k_2+k_3+k_4+k'_1+k'_2)}\big(n - (k_1 + k_2 + k_3 + k_4)\big)\nn 
&{}^{{1\over2}}C_{k_1}{}^{-{1\over2}}C_{k_2}{}^{{1\over2}}C_{k_3}{}^{-{1\over2}}C_{k_4}{}^{-{1\over2}}C_{k'_1}{}^{{1\over2}}C_{k'_2}{}^{-{1\over2}}C_{k'_3}{}^{{1\over2}}C_{m+n-(k_1+k'_1+k_2+k'_2+k_3+k'_3+k_4)}\nn 
&\times\Big(z^{k_1+k'_1}_{1}z^{k_2+k'_2}_{2}z^{k_3+k'_3}_{3}z^{m+n-(k_1+k'_1+k_2+k'_2+k_3+k'_3)}_{4}\nn
&\qquad + z^{k_1+k'_1}_{2}z^{k_2+k'_2}_{1}z^{k_3+k'_3}_{4}z^{m+n-(k_1+k'_1+k_2+k'_2+k_3+k'_3)}_{3}\Big) 
\nn [3pt]
& \! - A(x,\bar x)( z_1- z_3)( z_2 - z_4)\nn 
&\quad\times\bigg({1\over \sqrt2m}(-1)^{m}\sum_{k'_1=0}^{m-1}~\sum_{k'_2=0}^{m-(k'_1+1)}~\sum_{k'_3=0}^{m-(k'_1+k'_2+1)}{}^{-{1\over2}}C_{k'_1}{}^{-{1\over2}}C_{k'_2}{}^{-{1\over2}}C_{k'_3}{}^{-{1\over2}}C_{m-(k'_1+k'_2+k'_3+1)}
\nn
&\qquad\quad z_1^{k'_1}z_2^{k'_2}z_3^{k'_3}z_4^{m-(k'_1+k'_2+k'_3+1)}\bigg)
\nn
&\quad\times\bigg({1\over \sqrt2n}(-1)^{n}\sum_{k_1=0}^{n-1}~\sum_{k_2=0}^{n-(k_1+1)}~\sum_{k_3=0}^{n-(k_1+k_2+1)}{}^{-{1\over2}}C_{k_1}{}^{-{1\over2}}C_{k_2}{}^{-{1\over2}}C_{k_3}{}^{-{1\over2}}C_{n-(k_1+k_2+k_3+1)}
\nn
&\qquad\quad z_1^{k_1}z_2^{k_2}z_3^{k_3}z_4^{n-(k_1+k_2+k_3+1)}\bigg) 
\end{align}
The left-right mixed pair creation coefficient is
\begin{align}
\beta_{m,n} &= B(x,\bar x)|z_1-z_3||z_2-z_4|
\nn
&\quad\times\bigg({1\over m}(-1)^{m}\sum_{k'_1=0}^{m-1}~\sum_{k'_2=0}^{m-(k'_1+1)}~\sum_{k'_3=0}^{m-(k'_1+k'_2+1)}{}^{-{1\over2}}C_{k'_1}{}^{-{1\over2}}C_{k'_2}{}^{-{1\over2}}C'_{k_3}{}^{-{1\over2}}C_{m-(k'_1+k'_2+k'_3+1)}
\nn
&\quad\qquad~~ z_1^{k'_1}z_2^{k'_2}z_3^{k'_3}z_4^{m-(k'_1+k'_2+k'_3+1)}\bigg)
\nn 
&\quad\times\bigg({1\over n}(-1)^{n}\sum_{k_1=0}^{n-1}~\sum_{k_2=0}^{n-(k_1+1)}~\sum_{k_3=0}^{n-(k_1+k_2+1)}{}^{-{1\over2}}C_{k_1}{}^{-{1\over2}}C_{k_2}{}^{-{1\over2}}C_{k_3}{}^{-{1\over2}}C_{n-(k_1+k_2+k_3+1)}
\nn
&\quad\qquad~~ \bar z_1^{k_1}\bar z_2^{k_2}\bar z_3^{k_3}\bar z_4^{n-(k_1+k_2+k_3+1)}\bigg)
\end{align}
The right-moving pair creation coefficient is the complex conjugate of the left-moving pair creation coefficient
\begin{align} \label{gamma final prime}
\bar \gamma_{m,n} 
=\ & {1\over 4mn}(-1)^{m+n}\sum_{k_1=0}^{n-1}~\sum_{k_2=0}^{n-(k_1+1)}~\sum_{k_3=0}^{n-(k_1+k_2+1)}~\sum_{k_4=0}^{n-(k_1+k_2+k_3+1)}~\sum_{k'_1=0}^{m+n - (k_1+k_2+k_3+k_4)}\nn 
&~\sum_{k'_2=0}^{m+n - (k_1+k_2+k_3+k_4+k'_1)}\sum_{k'_3=0}^{m+n - (k_1+k_2+k_3+k_4+k'_1+k'_2)}\big(n - (k_1 + k_2 + k_3 + k_4)\big)\nn 
&{}^{{1\over2}}C_{k_1}{}^{-{1\over2}}C_{k_2}{}^{{1\over2}}C_{k_3}{}^{-{1\over2}}C_{k_4}{}^{-{1\over2}}C_{k'_1}{}^{{1\over2}}C_{k'_2}{}^{-{1\over2}}C_{k'_3}{}^{{1\over2}}C_{m+n-(k_1+k'_1+k_2+k'_2+k_3+k'_3+k_4)}\nn 
&\times\Big(\bar z^{k_1+k'_1}_{1}\bar z^{k_2+k'_2}_{2}\bar z^{k_3+k'_3}_{3}\bar z^{m+n-(k_1+k'_1+k_2+k'_2+k_3+k'_3)}_{4}\nn
&\qquad + \bar z^{k_1+k'_1}_{2}\bar z^{k_2+k'_2}_{1} \bar z^{k_3+k'_3}_{4}\bar z^{m+n-(k_1+k'_1+k_2+k'_2+k_3+k'_3)}_{3}\Big) 
\nn [3pt]
& \! - \bar A(\bar x, x)( \bar z_1- \bar z_3)(  \bar z_2 - \bar z_4)\nn 
&\quad\times\bigg({1\over \sqrt2m}(-1)^{m}\sum_{k'_1=0}^{m-1}~\sum_{k'_2=0}^{m-(k'_1+1)}~\sum_{k'_3=0}^{m-(k'_1+k'_2+1)}{}^{-{1\over2}}C_{k'_1}{}^{-{1\over2}}C_{k'_2}{}^{-{1\over2}}C_{k'_3}{}^{-{1\over2}}C_{m-(k'_1+k'_2+k'_3+1)}
\nn
&\qquad\quad \bar z_1^{k'_1}\bar z_2^{k'_2} \bar z_3^{k'_3}\bar z_4^{m-(k'_1+k'_2+k'_3+1)}\bigg)
\nn
&\quad\times\bigg({1\over \sqrt2n}(-1)^{n}\sum_{k_1=0}^{n-1}~\sum_{k_2=0}^{n-(k_1+1)}~\sum_{k_3=0}^{n-(k_1+k_2+1)}{}^{-{1\over2}}C_{k_1}{}^{-{1\over2}}C_{k_2}{}^{-{1\over2}}C_{k_3}{}^{-{1\over2}}C_{n-(k_1+k_2+k_3+1)}
\nn
&\qquad\quad \bar z_1^{k_1}\bar z_2^{k_2}\bar z_3^{k_3}\bar z_4^{n-(k_1+k_2+k_3+1)}\bigg) 
\end{align}
where the absolute value is given by $|z|=\sqrt{z\bar z}$, and the functions $A$ and $B$ and are defined as
\begin{align}
A(x,\bar x) &= {(E(x) - (1-x)K(x))\bar K(1-\bar x)-( E(1-x) - xK(1-x) )\bar K(\bar x) \over2 (K(x)\bar K(1-\bar x)+  K(1-x)\bar K(\bar x))}
\nn 
\nn
B(x,\bar x) &=\bar B(\bar x, x)={\pi\over 4(K(x)\bar K(1-\bar x)+  K(1-x)\bar K(\bar x))}
\nn
\nn
\bar A(\bar x, x) &= {(\bar E(\bar x) - (1-\bar x)\bar K(\bar x)) K(1- x)-( \bar E(1-\bar x) - \bar x\bar K(1-\bar x) )K(x) \over2 (K(x)\bar K(1-\bar x)+  K(1-x)\bar K(\bar x))}
\end{align}
Here $K(y)$ and $E(y)$ are the complete elliptic integrals of the first and second kinds respectively.
The cross ratio is defined as
\begin{align} 
x&={(z_1-z_2)(z_3-z_4)\over(z_1-z_3)(z_2-z_4)},\quad
\bar x={(\bar z_1-\bar z_2)(\bar z_3-\bar z_4)\over(\bar z_1-\bar z_3)(\bar z_2-\bar z_4)},\quad z_j=e^{w_j}, \quad \bar z_j=e^{\bar w_j}
\end{align}
where $j = 1,2,3,4$.
In Euclidean signature, the cylinder coordinates $w_j$ are given by
\begin{align} 
w_j &= \tau_j + i\sigma_j,\quad \bar w_j=\tau_j - i\sigma_j
\end{align}
while in Lorentzian signature
\begin{align} 
w_j &= i(t_j + \sigma_j),\quad \bar w_j =i(t_j -  \sigma_j)
\end{align}

\section{Discussion and Conclusion}\label{discussion}

In this paper, we investigate the pair creation effect of four twist operators. We focus on the simplest setup in which these operators act on the untwisted vacuum of two copies of a free boson.
After applying four twist operators, the state returns to the untwisted sector of two copies of the free boson and the resulting state is not the vacuum but rather a state with pairs of excitations. In this process, only antisymmetric modes (\ref{symm antisymm modes}) between the two copies can be created.

To compute this effect, we use the covering map method. First, we compute the correlation function of four twist operators and two bosons, following the derivation in \cite{Dixon:1986qv}. Next, we use the SL$(2,\mathbb C)$ symmetry to place the twist operators at arbitrary positions. Finally, we perform a mode expansion to derive the pair creation coefficients. Our results demonstrate that these coefficients exhibit complete symmetry with respect to the positions of the four twist operators, as expected, given that all twist-2 operators are identical. Additionally, we show that when any two twist operators coincide, the pair creation coefficients reduce to those for two twist operators, aligning with the interpretation of twist operators as operators of joining and splitting. In cases where a joining and splitting occur immediately at the same location, they leave no net effect on the state.

We identify three coefficients that capture the pair creation effect of four twist operators: $\gamma_{m,n}(z_i, \bar z_i)$, corresponding to a pair of left-moving modes; $\bar \gamma_{m,n} (z_i, \bar z_i)$, the conjugate of $\gamma_{m,n}$, corresponding to a pair of right-moving modes; and $\beta_{m,n}(z_i, \bar z_i)$, corresponding to a mixed pair with one left- and one right-moving mode. Compared to previous studies involving one or two twist operators, this result introduces a novel left-right coupling in two ways: first, these coefficients depend on both $z_i$ and $\bar z_i$, rather than being purely holomorphic or anti-holomorphic; and second, the presence of the new coefficient $\beta_{m,n}$ enables the creation of a mixed pair of one left and one right mover. We show that pair creation with these new features also satisfies the Bogoliubov ansatz (see Appendix \ref{proof}), which follows naturally from Wick contractions on the covering surface, even though in this case the covering surface is a torus. We find that, in general, mixed pair creation is at the same order as unmixed pair creation. We have also identified specific configurations in which mixed pair creation is dominant (see section \ref{sec numeri}).

These new features of left-right coupling originate from the torus covering surface, where the left- and right-moving modes are coupled due to the periodicities of the torus. This coupling on the torus becomes a coupling between left and right movers in the original base space. 
These periodicities also appear in the base space as monodromy conditions, which couple the left and right movers together. In recent discussions of the duality between tensionless strings and the free symmetric orbifold CFT, the covering surface is seen as dual to the string worldsheet \cite{Eberhardt:2018ouy,Eberhardt:2019ywk}. In this context, the left and right movers are coupled via the string loop in the bulk.

This form of left-right coupling differs from previous forms observed in the symmetric orbifold CFT. In the D1D5 CFT, a deformation is constructed using the same twist-2 operators, but dressed with both left- and right-moving supercharges, which couples left and right movers. Additionally, a physical state condition in the symmetric orbifold CFT requires that the left and right dimensions differ by an integer. In twisted sectors, where modes numbers and hence the dimensions of each mode can be fractional, this condition imposes specific constraints on the combinations of left and right movers that form a physical state. Both of these forms of left-right coupling, whether through supercharge dressing or the physical state condition, are noted in the literature. Our findings, however, reveal a distinct mechanism of direct left-right coupling arising purely through the twist operators.

This left-right coupling enables interesting processes, such as transitions from the untwisted vacuum to states with a single left-moving and right-moving mode. These processes cannot be produced by supercharge dressing or the physical state condition, as both introduce even numbers of left and right movers. In this paper, we have focused on the effect of four twist operators on the vacuum, but a similar analysis could be extended to excited states, introducing additional effects. For example, a left-moving mode can switch to a right-moving mode after passing through the four twist operators, or an initial pair of left- and right-moving modes could annihilate to produce the vacuum. We plan to address these effects in future work. Another direction would be to explore these new features in the context of thermalization \cite{Hampton:2019csz} and scrambling \cite{Guo:2021ybz,Chen:2024oqv} in the D1D5 CFT. Starting with high-energy initial modes, their interactions could result in a cascade of low-energy modes. Investigating the role of left-right coupling in thermalization could provide insights into these dynamics.

It would be interesting to extend the results obtained in this paper to scenarios with an arbitrary number of twist operators \cite{Atick:1987kd}. Adding more twist operators increases the genus of the covering surface, which corresponds to additional monodromy conditions. On a higher-genus surface, the left- and right-moving modes generally become coupled, leading to a left-right coupling effect in the presence of multiple twist operators (more than two). The effect of three twist operators, for instance, effectively creates a four-twist setup because it involves a twist operator insertion at future infinity, thus inherently requiring four twist insertions. For work on correlation functions on arbitrary Riemann surfaces, see \cite{Martinec:1986bq,Bonini:1987sk}. Understanding the general effect of multi-twist operators would be a major step forward in studying high-order perturbations in the D1D5 CFT, which may lead to a better understanding of the holographic duality. We hope to explore these ideas in future work.

\section*{Acknowledgements}
\appendix 
We would like to thank Emil Martinec and Samir Mathur for helpful discussions and suggestions. SDH would like to thank the hospitality of both APCTP in PoHang and KAIST in Daejeon during the completion of this work. The work of BG is supported by the ERC Grant 787320 - QBH Structure, the Simons Foundation grant 994306 (Simons Collaboration on Confinement and QCD Strings), and the NCCR SwissMAP funded by the Swiss National Science Foundation. The work of SDH is supported by KIAS Grant PG096301.

\section{Deriving the $z$-plane correlation functions}\label{mapping to z plane}\label{t to z}

In this appendix we show the derivation of the expressions for the $z$-plane correlation functions $g$ and $b$ at special points, recorded in (\ref{g b sp pts}). 
We first show the derivation of the constants $c_1$ and $c_2$ in the torus correlation function (\ref{c1 c2}).
We then show how to write the $z$-plane correlation function (\ref{g and b z to t prime}) completely in $z$-plane coordinates to obtain (\ref{g b sp pts}).
For various identities involving elliptic functions in the forthcoming computations we've utilized a variety of references \cite{Dixon:1986qv,Pastras:2017wot,Taylor:ellipticfns,functions2}. 

\subsection{Deriving constants $c_1$ and $c_2$}\label{der c1 c2}

To derive the constants $c_1$ and $c_2$ we first record their expressions below from (\ref{c1 c2})
\begin{align}\label{c1 c2 app}
c_1&={1\over2i\text{Im}\tau}\bigg(\int_0^{\tau}dt \mathcal P(t)-\bar\tau\int_0^1 dt\mathcal P(t)\bigg)
\nn  
c_2 &= -{1\over i\text{Im}\tau}\bigg(\int_0^{\tau} dt \mathcal P(t)-\tau\int_0^1 dt \mathcal P(t)\bigg)
\end{align}
We note the following relation
\bea 
\int^t_0 dt'\mathcal  P(t') = - \zeta(t) 
\eea
with $\zeta(t)$ being the Weierstrass $\zeta$ function (not to be confused with the Riemman zeta function) which is quasiperiodic under shifts along periods $\omega_i$, \cite{Dixon:1986qv,Pastras:2017wot,functions} with $\omega_i = (1,\tau)$ so 
\begin{equation}\label{quasi period}
\zeta(t+\omega_i) = \zeta(t) +2\eta_i
\end{equation}
where 
\bea\label{eta zeta rel}
\eta_i\equiv\zeta\Big({\omega_i\over2}\Big)
\eea
We note that $\zeta(t)$, has a single pole since the $\zeta(t)$ is proportional to the integral of $\mathcal P$ which has a double pole. It's form is
\bea 
\zeta(t)={1\over t} + \sum_{(m,n)\in\mathbb{Z}^2\backslash \{(0,0)\}}\bigg({1\over t+m\omega_1+n\omega_2}-{1\over m\omega_1+ n\omega_2}+ {t\over( m\omega_1+ n\omega_2)^2}\bigg)
\eea
Integrating $\zeta(t)$ around the boundary of the fundamental domain of the torus we have
\bea \label{perim}
\int_0^{1} dt\zeta(t) + \int_1^{1+\tau}dt\zeta(t) + \int_{1+\tau}^{\tau}\zeta(t) + \int_{\tau}^0\zeta(t)=\oint dt\zeta(t) = 2\pi i
\eea
Shifting integration regions for the second and third terms on the left hand side and using quasi-periodicity (\ref{quasi period}), we obtain
\bea\label{perim four terms}
\oint dt\zeta(t)=2\eta_1\tau -2\eta_2
\eea
Applying (\ref{perim}), we find the relation
\bea\label{eta1eta2}
\eta_1\tau - \eta_2=i\pi
\eea
which can be written as 
\bea \label{integral form}
-\tau\int_0^{1\over2} dt\mathcal P(t) + \int_0^{\tau\over2}dt\mathcal P(t)=i\pi
\eea
By using the parity and periodicity properties of $\mathcal P$
\bea 
\mathcal P(-t)=\mathcal P(t),\quad \mathcal P(t+\omega_i)=\mathcal P(t)
\eea
the relation (\ref{integral form}) can be transformed into the following
\bea \label{integral form prime}
-\tau\int_{0}^{1}dt\mathcal P(t) + \int_{0}^{\tau}dt\mathcal P(t)=2 \pi i
\eea
Using this to relate the integrals in (\ref{c1 c2 app}), we obtain expressions for the coefficients given by (\ref{c1 c2})
\begin{align}\label{c1 c2 prime}
c_1&=\int_0^1dt \mathcal P(t) +{\pi\over\text{Im}\tau}
\nn 
c_2 &= -{2\pi\over\text{Im}\tau}
\end{align}

\subsection{$z$-plane correlation functions at special points}

Here we derive the $z$-plane correlation functions $g$ and $b$ at special points, (\ref{g b sp pts}) and (\ref{AB}), from expression (\ref{g and b z to t prime}).
\subsubsection{Deriving $g$}
Starting with relation (\ref{g and b z to t prime}) 
we have
\begin{equation}\label{g and b z to t prime append 3}
g(z,z';x,\bar x)=\bigg({dz\over dt}\bigg)^{-1}\bigg({dz'\over dt'}\bigg)^{-1}\bigg(\mathcal{P}(t-t') + \mathcal{P}(t+t') - 2\int_0^1dt\mathcal{P}(t)-{2\pi\over\text{Im}\tau}\bigg)
\end{equation}
Let's note some helpful relations involving the Weierstrass $\mathcal P$ function, which will allow us to rewrite the above correlation function. We recall the map
\bea \label{t plane 2}
z = {\mathcal P(t) - e_1\over e_2 -e_1}
\eea
and relations
\begin{align} \label{half periods 2}
&e_1= \mathcal P\big({1\over2}\big),\quad e_2= \mathcal P\big({\tau\over2}\big),\quad e_3 = \mathcal P\big({1\over2}+{\tau\over2}\big),\quad e_1 + e_2 + e_3 = 0,\nn
& z\big({1\over2}\big)=0,\quad z\big({\tau\over2}\big)=1,\quad z\big({1\over2}+{\tau\over2}\big)=x,\quad z(0)=\infty,\quad x = {e_3-e_1\over e_2-e_1}
\end{align}
The addition formula for the Weierstrass $\mathcal P$ function is given by
\bea \label{add f}
\mathcal P(z+w) = {1\over4}\Big({\mathcal P'(z) - \mathcal P'(w)\over\mathcal P(z) - \mathcal P(w)}\Big)^2 - \mathcal P(z) -\mathcal P(w)
\eea
where $\mathcal P'(t)$ denotes the derivative of $\mathcal P(t)$ with respect to the argument. We also have the following identity
\bea
\mathcal P'^2(t) = 4(\mathcal P(t)-e_1)(\mathcal P(t)-e_2)(\mathcal P(t)-e_3)
\eea
which, using (\ref{t plane 2}) and (\ref{half periods 2}), yields
\bea\label{der of P}
 \mathcal P'(t)=2z^{1\over2}(z-1)^{1\over2}(z-x)^{1\over2}(e_2-e_1)^{3\over2}
\eea
Substituting
\be \label{dzdt}
{dz\over dt} ={1\over e_2-e_1}\mathcal P'(t)
\ee
and the addition formula (\ref{add f}) into (\ref{g and b z to t prime append 3}) yields
\begin{align}
& g(z,z';x,\bar x)\nn
=\ &  (e_2-e_1)^2{1\over2}{\mathcal P'(t)\over\mathcal{P}'(t')}{1 \over(\mathcal P(t) - \mathcal P(t'))^2}
+(e_2-e_1)^2{1\over2}{ \mathcal P'(t')\over\mathcal{P}'(t)}{1\over(\mathcal P(t) - \mathcal P(t'))^2}\nn
& - 2(e_2-e_1)^2{\mathcal P(t) +\mathcal P(t')\over\mathcal{P}'(t)\mathcal{P}'(t')} - 2(e_2-e_1)^2{1\over\mathcal{P}'(t)\mathcal{P}'(t')} \bigg(\int_0^1dt\mathcal{P}(t)+{\pi\over\text{Im}\tau}\bigg)
\end{align}
Using (\ref{t plane 2}) and (\ref{der of P}), the above correlation function can be written as
\begin{align}
& g(z,z';x,\bar x)\nn
=\ &  {1\over2}{z^{1\over2}(z-1)^{1\over2}(z-x)^{1\over2}\over z'^{{1\over2}}(z'-1)^{1\over2}(z'-x)^{1\over2}}{1 \over(z-z')^2}
+{1\over2}{z'^{{1\over2}}(z'-1)^{1\over2}(z'-x)^{1\over2}\over z^{1\over2}(z-1)^{1\over2}(z-x)^{1\over2}}{1\over (z-z')^2}\nn
& -  {1\over2}{z+z'\over z^{1\over2}(z-1)^{1\over2}(z-x)^{1\over2}z'^{1\over2}(z'-1)^{1\over2}(z'-x)^{1\over2}}\nn  
&-{1\over z^{1\over2}(z-1)^{1\over2}(z-x)^{1\over2}z'^{{1\over2}}(z'-1)^{1\over2}(z'-x)^{1\over2}}  {1\over2(e_2-e_1)}\bigg(2e_1+\int_0^1dt\mathcal{P}(t)+{\pi\over\text{Im}\tau}\bigg)
\end{align}
which can be rewritten as
\begin{align}\label{gj}
&g(z,z';x,\bar x)\nn 
=\ &  {1\over2}{z^{1\over2}(z-1)^{1\over2}(z'-x)^{1\over2}\over z'^{{1\over2}}(z'-1)^{1\over2}(z-x)^{1\over2}}{1\over(z-z')^2}
+ {1\over2}{z'^{{1\over2}}(z'-1)^{1\over2}(z-x)^{1\over2}\over z^{{1\over2}}(z-1)^{1\over2}(z'-x)^{1\over2}}{1\over(z-z')^2}
\nn
&-{1\over z^{1\over2}(z-1)^{1\over2}(z-x)^{1\over2}z'^{{1\over2}}(z'-1)^{1\over2}(z'-x)^{1\over2}} \bigg[ {1\over2(e_2-e_1)}\bigg(-e_3 +\int_0^1dt\mathcal{P}(t)+{\pi\over\text{Im}\tau}\bigg)\bigg]\nn
\end{align}
We note the relation which expresses the elliptic zeta function as a sum
\be\label{intP}
\int_0^1dt\mathcal P(t)=-\zeta(1)=(2\pi i )^2\Big({1\over12} - 2\sum_{n=1}^{\infty}{nu^n\over 1-u^n}\Big),\qquad u = e^{2\pi i \tau}
\ee
which allows us to write
\be
\int_0^1dt\mathcal P(t)= (2\pi i){d\over d\tau}\big(\ln u^{1\over12}\prod_{n=1}^{\infty}(1-u^n)^2\big)
=(2\pi i){d\over d\tau}\ln \eta(\tau)^2
\ee
where we have used the $\eta$ function
\be
\eta(\tau)=u^{1\over24}\prod_{n=1}^{\infty}(1-u^n)
\ee
Using the relation between the $\eta$ function and the theta functions, $\vartheta_i(\tau)$ for $i=1,2,3$
\bea 
 \eta(\tau)^3=\frac{1}{2}\vartheta_2(\tau)\vartheta_3(\tau)\vartheta_4(\tau)
\eea
and an additional theta function relation
\bea\label{theta234 relation}
\vartheta^4_2(\tau)=\vartheta_3^4(\tau)-\vartheta_4^4(\tau)
\eea
we find that the integral over the Weierstrass function can be written as
\be \label{int P}
\int_0^1dt\mathcal P(t)={\pi i\over 3}{d\over d\tau}\ln \big(\vartheta_2(\tau)\vartheta_3(\tau)\vartheta_4(\tau) \big)^{4}
={\pi i \over 3}{d\over d\tau}\ln{(\vartheta_3^4(\tau)-\vartheta_4^4(\tau))(\vartheta_3(\tau)\vartheta_4(\tau))^4}
\ee
Recalling the cross ratio $x$, (\ref{cross ratio}), written in terms of $\vartheta_3(\tau)$ and $\vartheta_4(\tau)$ and writing an additional relation that relates $\vartheta_3(\tau)$ with Hypergeometric functions which are functions of the cross ratio, $x$, 
\begin{align}\label{F}
F(x)&\equiv {}_2F_1\bigg({1\over2},{1\over2};1;x\bigg)={1\over \pi }\int_0^1 dy y^{-{1\over2}}(1-y)^{-{1\over2}}(1-xy)^{-{1\over2}}
\nn 
F(1-x)&= {}_2F_1\bigg({1\over2},{1\over2};1;1-x\bigg)
\end{align}
we have
\begin{align}\label{more relations}
x = {\vartheta^4_4(\tau)\over\vartheta^4_3(\tau)},\quad\vartheta^4_3(\tau) = F(1-x)^2,\quad \vartheta^4_4(\tau) = xF(1-x)^2,\quad F(x)=\vartheta^2_3(-1/\tau)=-i\tau\vartheta^2_3(\tau)\nn 
\end{align}
With the above identities, (\ref{int P}) can be written as
\begin{align} \label{int P 2}
\int_0^1dt\mathcal P(t)&={\pi i\over 3}{dx\over d\tau}{d\over dx}\ln \big(x(1-x)F(1-x)^6\big)
\end{align}
We note the additional relations between the $e_i$'s and the $\vartheta_i(\tau)$'s
\begin{align}
e_1 &= {\pi^2\over3}(\vartheta^4_3(\tau)+\vartheta^4_4(\tau))\nn 
e_2 &=-{\pi^2\over3}(\vartheta^4_2(\tau)+\vartheta^4_3(\tau))\nn 
e_3 &= {\pi^2\over3}(\vartheta^4_2(\tau)-\vartheta^4_4(\tau))
\end{align}
which have convenient difference relations
\begin{align}\label{e diff relations}
e_2 - e_1 &= -\pi^2\vartheta_3^4(\tau)\nn
e_2 - e_3 &=-\pi^2\vartheta_2^4(\tau)\nn
e_3 - e_1&= -\pi^2\vartheta_4^4(\tau)
\end{align} 
Using the above relations, the $e_i$'s can be expressed in terms of hypergeometric functions.
To compute the third term in (\ref{gj}), we will need
\begin{align}\label{e3 prime}
e_3&={\pi^2\over3}(\vartheta_3^4(\tau) - 2\vartheta_4^4(\tau) )={\pi^2\over3}F(1-x)^2\big(1-2x\big)\nn
e_2 - e_1 &= -\pi^2\vartheta_3^4(\tau)= -\pi^2 F(1-x)^2
\end{align} 
Using (\ref{more relations}) we note that $\tau$ and $x$ are related in the following way 
\bea\label{tau taubar}
\tau(x) = {iF(x)\over F(1-x)},\quad \bar\tau(\bar x) = -{i\bar F(\bar x)\over \bar F(1-\bar x)}
\eea
which gives
\bea \label{im tau}
 \text{Im} \tau = {F(x)\bar F(1-\bar x)+F(1-x)\bar F(\bar x)\over 2F(1-x)\bar F(1-\bar x)}
\eea
A set of useful relations between hypergeometric functions with elliptic functions are given by
\begin{align}\label{F dF}
    F(x) &= {2 \over\pi} K(x),\qquad \qquad \quad{d\over dx} F(x)= {E(x) + (x-1)K(x)\over\pi x(1-x)}\nn
    F(1-x)&={2 \over\pi} K(1-x),\qquad {d\over dx} F(1-x)= {E(1-x) - xK(1-x)\over\pi x(x-1)}
\end{align}
where $K(x)$ and $E(x)$ are complete elliptic integrals of the first and second kinds respectively. Therefore using (\ref{F dF}) and (\ref{tau taubar}) the derivative of ${d\tau\over dx}$ can be computed as
\begin{align}
 {d\tau\over dx} &= {i\over F(1-x)}\Big( {d F(x)\over dx} - {F(x)\over F(1-x)}{d\over dx}F(1-x) \Big) \nn
 &={2i\over\pi^2 x(1-x) F(1-x)^2} \Big( E(x) K(1-x) + K(x)E(1-x) -K(x) K(1-x) \Big)
\end{align}
Utilizing the identity
\begin{equation}
    E(x) K(1-x) + K(x)E(1-x) -K(x) K(1-x) = {\pi\over2}
\end{equation}
we find the expression
\begin{equation}\label{dtaudx}
  {d\tau\over dx} =   {i\over\pi x(1-x) F(1-x)^2} 
\end{equation}
To compute the following constant in the third term in (\ref{gj}), we collect together (\ref{e3 prime}), (\ref{int P 2}), (\ref{dtaudx}) and (\ref{im tau}), which yields
\begin{align}\label{constant 2}
&-{1\over2(e_2-e_1)}\bigg(-e_3+\int_0^1dt\mathcal{P}(t)+{\pi\over\text{Im}\tau}\bigg)
\nn 
=\ & x(1-x) {d\over dx}\ln\big(F(x)\bar F(1-\bar x)+F(1-x)\bar F(\bar x)\big)\equiv A(x,\bar x)
\end{align}
where for the third term in parenthesis, ${\pi\over \text{Im}\tau}$, it helps to utilize the rewriting
\begin{equation}
    {\pi\over \text{Im}\tau} = \pi{dx\over d\tau}{1\over \text{Im}\tau}{d\tau\over dx} = 2\pi i{dx\over d\tau}{d\over dx}\ln \text{Im}\tau
\end{equation}
Inserting (\ref{constant 2}) into the correlation function (\ref{gj}) yields
\begin{align}\label{gjfinal1}
g(z,z';x,\bar x)
&=\, {{1\over2}z(z-1)(z'-x) + {1\over2}z'(z'-1)(z-x)+(z-z')^2 A(x,\bar x)\over z^{1\over2}(z-x)^{1\over2}(z-1)^{1\over2}z'^{{1\over2}}(z'-x)^{1\over2}(z'-1)^{1\over2}(z-z')^2}
\end{align}
This result agrees with the eq. (4.22) in \cite{Dixon:1986qv} for $N=2$, which is expressed for general twist locations but $A$ (as given in (4.30)) is computed for special twist locations. That result was obtained directly in the $z$-plane using monodromy conditions.

\subsubsection{Deriving $b$}

Starting with relation (\ref{g and b z to t prime}) we have
\be
b(\bar z,z';x,\bar x)=\bigg({d\bar z\over d\bar t}\bigg)^{-1}\bigg({d z'\over d t'}\bigg)^{-1}{2\pi\over\text{Im}\tau}
\ee
Utilizing (\ref{dzdt}) and (\ref{der of P}) we find that the correlation function becomes
\begin{align}\label{b z}
b(\bar z,z';x,\bar x)&= |e_2-e_1|^2{1\over\bar{\mathcal{P}}'(\bar t)}{1\over\mathcal{P}'(t')}{2\pi\over\,\text{Im}\tau}
\nn 
&={B(x,\bar x) \over\bar z^{1\over2}(\bar z - \bar x)^{1\over2}(\bar z - \bar 1)^{1\over2} z'^{{1\over2}}( z' - x)^{1\over2}( z' - 1)^{1\over2}}
\end{align}
where
\bea
B(x,\bar x) \equiv {1\over 2|e_2-e_1|}{\pi\over\text{Im} \tau} =  {1\over\pi (F(x)\bar F(1-\bar x)+F(1-x)\bar F(\bar x))} 
\eea
where we have used the relations (\ref{e3 prime}) and (\ref{im tau}).

\section{Proof of Bogoliubov ansatz}\label{proof}

In this appendix, we present a proof of the Bogoliubov ansatz as given in (\ref{ansatz 4 first}).
In \cite{Guo:2023czj},
the Bogoliubov ansatz was proven for any number of twist operators, though without mixing between left and right movers. Here, we generalize the proof to include left and right mover mixing, which occurs when the covering surface is not a sphere. The key idea is that the Bogoliubov ansatz originates from the Wick contraction on the covering surface. Even on the torus, where left and right mover contractions occur, the field remains free and continues to satisfy the structure of Wick contraction.
For clarity, we present the proof using four twist operators, and the method can be applied similarly to any number of twist operators.

Let us record the ansatz below
\begin{align}\label{ansatz 4 first a}
|\chi\rangle&\equiv\sigma(w_4,\bar w_4)\sigma(w_3,\bar w_3)\sigma(w_2,\bar w_2)\sigma(w_1,\bar w_1)|0\rangle\nn 
&=\mathcal C\exp\Big(\sum_{m,n>0}\gamma_{m,n}(w_i,\bar w_i)\a_{-m}\a_{-n} + \sum_{m,n>0}\beta_{m,n}(w_i,\bar w_i)\alpha_{-m}\bar{\alpha}_{-n}\nn
&\hspace{2cm}+\sum_{m,n>0}\bar\gamma_{m,n}(w_i,\bar w_i)\bar\a_{-m}\bar\a_{-n}\Big)|0\rangle
\end{align}
where 
\bea 
\mathcal C \equiv\langle 0| \sigma(w_4,\bar w_4)\sigma(w_3,\bar w_3)\sigma(w_2,\bar w_2)\sigma(w_1,\bar w_1)|0\rangle
\eea
From this ansatz, we have shown that the pair creation, $\gamma_{m,n}$, $\bar\gamma_{m,n}$ and $\beta_{m,n}$ as defined in (\ref{gamma11}), and (\ref{z plane aabar}) are given in the $z$-plane by
\begin{align} \label{pair creation z plane app}
\gamma_{m,n}&=-{1\over 2mn}\oint{dz\over2\pi }z^n{dz'\over2\pi }z'^m g(z,z';z_i,\bar z_i)
\nn
\beta_{m,n} &= -{1\over  m n}\oint{d\bar z\over2\pi }\bar z^{n}{dz'\over2\pi }z'^m\, b(\bar z, z';z_i,\bar z_i)
\end{align} 
Here and throughout this appendix, the $\oint$ means integration over contours where $|z|,|z'|>|z_i|$. The corresponding expression for $\bar\gamma_{m,n}$ is analogous.
These quantities can be computed by first evaluating the correlation functions that involve two insertions of bosonic fields and four twist operators, as defined in (\ref{g})
\begin{align}\label{g appendix}
g(z,z';z_i,\bar z_i) &\equiv-{\langle 0|\partial_z X\partial_{z'} X \prod_{i=1}^4\sigma(z_i, \bar z_i) |0\rangle\over \langle 0| \prod_{i=1}^4\sigma(z_i, \bar z_i)|0\rangle}
\nn
b(\bar z, z';z_i,\bar z_i)&\equiv-{\langle0| \partial_{\bar z}X\partial_{ z'}X\prod_{i=1}^4\sigma(z_i, \bar z_i) |0\rangle\over \langle0| \prod_{i=1}^4\sigma(z_i, \bar z_i)|0\rangle}
\end{align}
To simplify the proof of the ansatz, let us make the following re-definitions 
\begin{align}\label{ansatz 4 first ccp}
|\chi\rangle=\mathcal C\exp\Big(\sum_{c,c'}\sum_{m,n>0}\gamma^{(c)(c')}_{m,n}\a^{(c)}_{-m}\a^{(c')}_{-n}\Big)|0\rangle
\end{align}
where the indices $c,c'=L,R$, and
\begin{align}
    \alpha^{(L)}_m&\equiv \alpha_m,\quad \alpha^{(R)}_m\equiv \bar\alpha_m
\end{align}
The pair creation coefficients are redefined as follows
\begin{equation}
\gamma^{(L)(L)}_{m,n}\equiv\gamma_{m,n},\quad \gamma^{(L)(R)}_{m,n}=\gamma^{(R)(L)}_{n,m}\equiv{\beta_{m,n}\over2},\quad \gamma^{(R)(R)}_{m,n}\equiv\bar\gamma_{m,n}
\end{equation}
We also define
\begin{equation}\label{wz redef}
w^{(L)}\equiv w,\quad w^{(R)}\equiv \bar w,\quad
z^{(L)} \equiv z =e^{w^{(L)}},\quad z^{(R)}\equiv\bar z=e^{w^{(R)}}
\end{equation}
With these, the pair creation coefficients can be written as
\begin{align} \label{pair creation z plane app pr}
\gamma^{(c)(c')}_{m,n}&\equiv-{1\over 2mn}\oint{dz^{(c)}\over2\pi }(z^{(c)})^n{dz'^{(c')}\over2\pi }(z'^{(c')})^mg^{(c)(c')}(z^{(c)},z'^{(c')})
\end{align}
where 
\begin{align}\label{gcc}
    g^{(L)(L)}(z^{(L)},z'^{(L)}) &\equiv g(z,z';z_i,\bar z_i)\nn 
     g^{(L)(R)}(z^{(L)},z'^{(R)}) &\equiv b( \bar z, z';z_i,\bar z_i)\nn
     g^{(R)(L)}(z^{(R)},z'^{(L)}) &\equiv b(\bar z, z';z_i,\bar z_i)\nn
     g^{(R)(R)}(z^{(R)},z'^{(R)}) &\equiv \bar g(\bar z,\bar z',\bar z_i, z_i)
\end{align}
For correlation functions involving an arbitrary even number of modes we can follow a similar procedure by mapping to the covering surface. Let's consider a correlation function with $2j$ number of modes acting on the final state, 
\begin{equation}\label{corr}
    {\langle0|\a^{(c_1)}_{m_{1}}\a^{(c_2)}_{m_{2}}\ldots \a^{(c_{2j})}_{m_{{2j}}}\prod_{i=1}^4\sigma(w_i, \bar w_i)|0\rangle\over\langle0|\prod_{i=1}^4\sigma(w_i, \bar w_i)|0\rangle}
\end{equation}
where $c_j=L,R$. Mapping to the $z$-plane and using the Bogoliubov ansatz (\ref{ansatz 4 first ccp}), we obtain
\begin{align}\label{RHS}
 {\langle0|\a^{(c_1)}_{m_{1}}\a^{(c_2)}_{m_{2}}\ldots \a^{(c_{2j})}_{m_{{2j}}}\prod_{i=1}^4\sigma(z_i, \bar z_i)|0\rangle\over\langle0|\prod_{i=1}^4\sigma(z_i, \bar z_i)|0\rangle}
={1\over j!}\big(\prod_{i=1}^{2j}m_i\big)\sum_{\mu}\prod^{2j}_{i=1,\,i\,\text{odd}}\gamma^{(c_{\mu(i)})(c_{\mu(i+1)})}_{m_{\mu(i)},m_{\mu(i+1)}}
\end{align}
where $\mu$ is a sum over all permutations of $2j$ indices. Now, let us
verify that this result from the Bogoliubov ansatz can be reproduced by directly computing the correlation function using Wick contractions on the covering surface. 
Utilizing the mode definitions in (\ref{bosonic modes z plane}), we write the left hand side of (\ref{RHS}) in the following way 
\begin{align}
&{\langle0|\a^{(c_1)}_{m_{1}}\a^{(c_2)}_{m_{2}}\ldots \a^{(c_{2j})}_{m_{{2j}}}\prod_{i=1}^4\sigma(z_i, \bar z_i)|0\rangle\over\langle0|\prod_{i=1}^4\sigma(z_i, \bar z_i)|0\rangle}\nn
\equiv \ & \oint\prod_{k=1}^{2j}{d z'^{(c_{k})}_{k}\over2\pi } (z'^{(c_{k})}_{k})^{m_{k}}
{\langle 0|\prod_{k=1}^{2j}\partial_{ z'^{(c_k)}_{k}}X \prod_{i=1}^4\sigma(z_i, \bar z_i)|0\rangle\over\langle 0|\prod_{i=1}^4\sigma(z_i, \bar z_i)|0\rangle}\nn
= \ & \oint\prod_{k=1}^{2j}{d \tilde z'^{(c_{k})}_{k}\over2\pi } (z'^{(c_{k})}_{k})^{m_{k}}
{\langle 0|\prod_{k=1}^{2j}\partial_{ \tilde{z}'^{(c_k)}_{k}}X \prod_{i=1}^4\sigma(\tilde z_i, \bar{\tilde {z_i}})|0\rangle\over\langle 0|\prod_{i=1}^4\sigma(\tilde z_i, \bar{\tilde {z_i}})|0\rangle}
\end{align}
In the last step, we apply an SL$(2,\mathbb C)$ transformation, $\tilde z(z)$, to map the positions of the twist operators $z_i$ to $\tilde z_1=0,\tilde z_2=x,\tilde z_3=1,\tilde z_4=\infty$.
By mapping the correlation function to the covering surface with coordinate $t$, using (\ref{def X}) and (\ref{t plane}), we obtain
\begin{align}\label{tplane}
&{\langle0|\a^{(c_1)}_{m_{1}}\a^{(c_2)}_{m_{2}}\ldots \a^{(c_{2j})}_{m_{{2j}}}\prod_{i=1}^4\sigma(z_i, \bar z_i)|0\rangle\over\langle0|\prod_{i=1}^4\sigma(z_i, \bar z_i)|0\rangle}\nn
= \ & \oint\prod_{k=1}^{2j}{d  t^{(c_{k})}_{k}\over2\pi } (z'^{(c_{k})}_{k})^{m_{k}}
\big\langle\prod_{k=1}^{2j}\frac{1}{\sqrt{2}}(\partial_{ t^{(c_k)}_{k}}X-\partial_{ -t^{(c_k)}_{k}}X) \big\rangle
\end{align}
where we have also redefined the $t$-surface coordinates as 
\begin{equation}
    t^{(L)}\equiv t,\quad t^{(R)}\equiv\bar t
\end{equation}
Now that we have mapped to the covering space we can freely perform pairwise Wick contractions of the bosonic fields. We have the following general Wick contraction term 
\begin{align}\label{gtau btau}
g^{(c)(c')}(t^{(c)},t'^{(c')})
&\equiv-\big\langle{1\over\sqrt2}(\partial_{t^{(c)}}X - \partial_{-t^{(c)}}X){1\over\sqrt2}(\partial_{t'^{(c')}}X - \partial_{-t'^{(c')}}X)\big\rangle
\end{align}
where the $g^{(c)(c')}(t^{(c)},t'^{(c')})$ are given by
\begin{align}\label{contr}
g^{(L)(L)}(t^{(L)},t'^{(L)})
&\equiv \mathcal{P}(t-t') + \mathcal{P}(t+t') - 2\int_0^1 dt\mathcal{P}(t)-{2\pi\over\text{Im}\tau}
\nn
g^{(R)(R)}(t^{(R)},t'^{(R)})
&\equiv \bar{\mathcal{P}}(\bar t-\bar t') + \bar{\mathcal{P}}(\bar t+\bar t') - 2\int_0^1d\bar t\bar{\mathcal{P}}(\bar t)-{2\pi\over\text{Im}\tau}
\nn
g^{(L)(R)}(t^{(L)},t'^{(R)})&=g^{(R)(L)}(t^{(R)},t'^{(L)})\equiv{2\pi\over\text{Im}\tau}
\end{align}
Performing Wick contractions of the bosonic fields in (\ref{tplane}), whose expressions are defined in (\ref{contr}), and mapping back to the $z$-plane we obtain
\begin{align}
&{\langle0|\a^{(c_1)}_{m_{1}}\a^{(c_2)}_{m_{2}}\ldots \a^{(c_{2j})}_{m_{{2j}}}\prod_{i=1}^4\sigma(z_i, \bar z_i)|0\rangle\over\langle0|\prod_{i=1}^4\sigma(z_i, \bar z_i)|0\rangle}\nn
= & \ {(-1)^j\over2^jj!}\oint\prod_{k=1}^{2j}{d  z'^{(c_{k})}_{k}\over2\pi } (z'^{(c_{k})}_{k})^{m_{k}}\sum_{\mu}\prod_{k=1,\,k\,\text{odd}}^{2j-1}g^{(c_{\mu(k)})(c_{\mu(k+1)})}(z'^{(c_{\mu(k)})}_{\mu(k)},z'^{(c_{\mu(k+1)})}_{\mu(k+1)})
\end{align}
where 
\begin{align}
&g^{(c)(c')}(z^{(c)},z'^{(c')})={d t^{(c)}\over d  z^{(c)}}
{d t'^{(c')}\over d  z'^{(c')}}
g^{(c)(c')}(t^{(c)},t'^{(c')})
\end{align}
which, utilizing (\ref{pair creation z plane app pr}), yields agreement with the right hand side of (\ref{RHS}), thereby verifying the Bogoliubov ansatz.

\bibliographystyle{JHEP}
\bibliography{bibliography.bib}

\end{document}